\documentclass[10pt,journal,compsoc]{IEEEtran}
\ifCLASSOPTIONcompsoc
  \usepackage[nocompress]{cite}
\else
  \usepackage{cite}
\fi
\ifCLASSINFOpdf
\else
\fi

\usepackage{makecell}
\usepackage{amsmath,amsfonts}
\usepackage{algorithmic}
\usepackage{algorithm}
\usepackage{array}
\usepackage{textcomp}
\usepackage{stfloats}
\usepackage{url}
\usepackage{color}
\usepackage{verbatim}
\usepackage{subcaption}
\usepackage{graphicx}
\usepackage{epstopdf}
\usepackage{cite}
\usepackage{multirow}
\usepackage{hyperref}
\usepackage{ragged2e}
\usepackage{subcaption}
\usepackage{float}
\begin{document}
%
\title{Continuous-variable Quantum Diffusion Model for State Generation and Restoration }

\author{Haitao~Huang,
        Chuangtao~Chen,
		Qinglin~Zhao
\thanks{Corresponding author: Qinglin Zhao.}
\IEEEcompsocitemizethanks{\IEEEcompsocthanksitem Haitao Huang, Chuangtao Chen and Qinglin Zhao are the Faculty of Innovation Engineering, Macau University of Science and Technology, Macao 999078, China. (e-mail: huanghaitao4396@gmail.com; chuangtaochen@gmail.com; qlzhao@must.edu.mo)}}

\IEEEtitleabstractindextext{%
\begin{abstract}
The generation and preservation of complex quantum states against environmental noise are paramount challenges in advancing continuous-variable (CV) quantum information processing. This paper introduces a novel framework based on continuous-variable quantum diffusion principles, synergizing them with CV quantum neural networks (CVQNNs) to address these dual challenges. For the task of state generation, our Continuous-Variable Quantum Diffusion Generative model (CVQD-G) employs a physically driven forward diffusion process using a thermal loss channel, which is then inverted by a learnable, parameter-efficient backward denoising process based on a CVQNN with time-embedding. This framework's capability is further extended for state recovery by the Continuous-Variable Quantum Diffusion Restoration model (CVQD-R), a specialized variant designed to restore quantum states, particularly coherent states with unknown parameters, from thermal degradation. Extensive numerical simulations validate these dual capabilities, demonstrating the high-fidelity generation of diverse Gaussian (coherent, squeezed) and non-Gaussian (Fock, cat) states, typically with fidelities exceeding 99\%, and confirming the model's ability to robustly restore corrupted states. Furthermore, a comprehensive complexity analysis reveals favorable training and inference costs, highlighting the framework's efficiency, scalability, and its potential as a robust tool for quantum state engineering and noise mitigation in realistic CV quantum systems.
\end{abstract}

}

\maketitle

\IEEEdisplaynontitleabstractindextext

%
\IEEEpeerreviewmaketitle

\IEEEraisesectionheading{\section{Introduction}\label{sec:introduction}}

%
%
%
%
\IEEEPARstart{Q}{uantum} engineering harnesses quantum mechanical principles to advance technologies in communication \cite{Liao2017quantumcommunication, gisin2007quantumcomm}, computing \cite{Ladd2010quantumcomputerqubit, Zhong2020quantumcomputerphoton}, and sensing \cite{Degen2017quantumsensing}. The processing of quantum information within these systems typically follows one of two primary paradigms: discrete-variable (DV) systems, which utilize qubits with finite-dimensional Hilbert spaces, and continuous-variable (CV) systems, which operate with observables possessing continuous spectra in infinite-dimensional Hilbert spaces. The latter forms the foundation for CV Quantum Computing, an approach that leverages these continuous degrees of freedom.

 CV Quantum Computing  presents several compelling advantages, including the potential for higher information encoding efficiency per physical system \cite{braunstein2005quantum}, inherent compatibility with existing optical fiber communication infrastructure \cite{Liao2017quantumcommunication}, and the capability to implement powerful non-Gaussian operations crucial for achieving universal quantum computation \cite{lloyd1999quantum}. A key application scenario that benefits from these features is quantum communication, which involves the preparation, transmission, and reception of quantum states encoded in optical modes (often termed qumodes). Environmental thermal noise poses a significant threat to quantum states during their transmission through physical channels like optical fibers or free space. For practical  CV Quantum Computing  systems, which heavily rely on such optical channels \cite{kraus1971general}, addressing the detrimental impact of this noise is therefore a critical challenge. This noise induces decoherence, transforming initially pure quantum states into mixed states and degrading the encoded quantum information \cite{Holevo_2012}. Such degradation compromises the accuracy of quantum computations and the efficiency of quantum communication. Therefore, the development of robust methods for both high-fidelity quantum state \textit{generation} (the initial preparation of desired states) and effective state \textit{restoration} (the recovery of states after noise-induced corruption) has become essential for the continued advancement and practical realization of  CV Quantum Computing .

Several quantum machine learning approaches have been explored for CV systems, including Continuous-Variable Quantum Neural Networks (CVQNNs) \cite{killoran2018continuous}, CV Boltzmann Machines \cite{bangar2024boltzmann}, CV Born Machines \cite{cepaite2022born}, and Optical GANs \cite{shrivastava2019optical}. While demonstrating capabilities in generating complex quantum states, these existing works often face limitations: 1) they primarily \textbf{focus on state generation tasks}, with less emphasis on the equally crucial problem of restoring states corrupted by noise, and 2) existing approaches often employ rigid frameworks relying on fixed transformations (e.g., specific unitary operations). Within these, gate parameters are typically trained to map a specific, predefined initial state (like a vacuum state) to a desired target output. Since the resulting transformation and its parameters are optimized solely for this fixed pathway, altering the initial state or target output generally requires model retraining. This inherent lack of adaptability makes them less suitable for dynamic scenarios such as quantum communication, where unpredictable noise demands a responsive processing strategy rather than a fixed one.

\subsection{Motivation}
\label{subsec:motivation}
The limitations of existing CV quantum machine learning models motivate the development of a new approach capable of addressing the dual challenge of \textbf{state generation and restoration} within a unified framework. Specifically, there is a need for:
\begin{enumerate}

    \item A \textbf{flexible and powerful framework} that can adapt to varying levels of noise, moving beyond the constraints of fixed input-output mappings inherent in many previous models. This adaptability is crucial for practical applications where quantum states undergo unpredictable degradation.
    \item Robust capabilities for high-fidelity quantum \textbf{state generation}, ideally starting from easily preparable initial states like thermal states.
    \item Effective quantum \textbf{state restoration} mechanisms that can take noise-corrupted states (even with unknown levels of pollution) as input and recover the original, clean quantum states with high fidelity. This directly addresses the gap left by generation-focused models.
\end{enumerate}
Meeting these requirements would significantly enhance the practicality and robustness of CV quantum technologies, particularly in communication and computation tasks operating in noisy environments.

\begin{figure}[t]
	\centering
	\includegraphics[width=\linewidth]{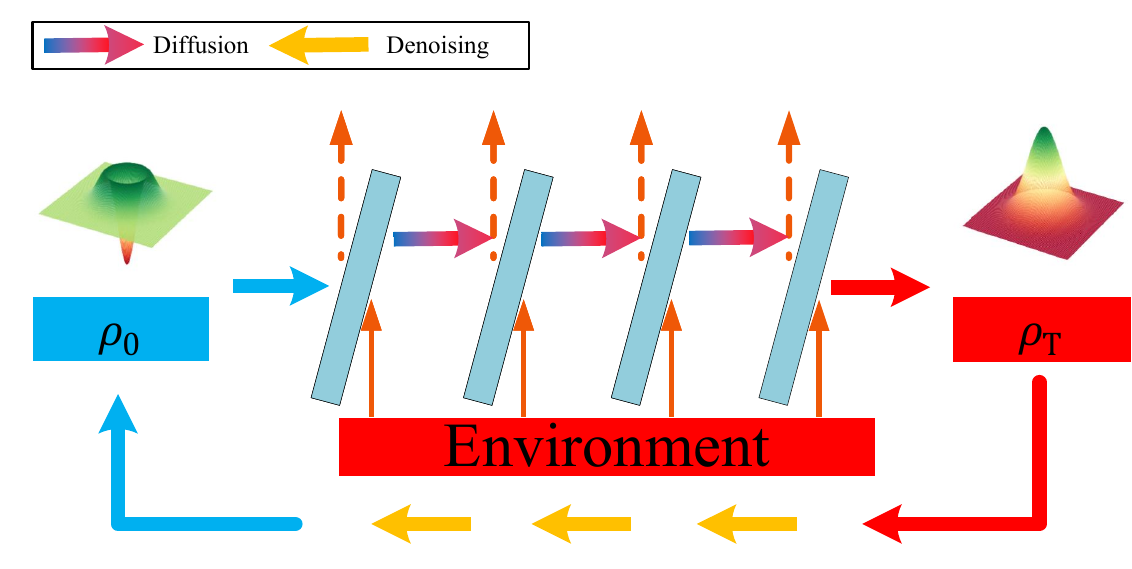} 
	\caption{The Continuous-Variable (CV) Quantum Diffusion Process: An Illustration via State Generation}
	\label{fig:main_idea}
\end{figure}

\begin{table*}[t]
\centering
\caption{Comparison Between CVQD-G and CVQD-R.}
\label{tab:model_comparison_new_pos}
\begin{tabular}{|p{0.25\textwidth}|p{0.35\textwidth}|p{0.35\textwidth}|}
\hline
\textbf{Feature} & \textbf{CVQD-G (Generative Model)} & \textbf{CVQD-R (Restoration Model)} \\
\hline
Forward Process & Start from a \textit{specific, predefined target state}. & Start from \textit{randomly sampled coherent states} (within a defined range). \\
\hline
Backward Process & A \textit{standard noise state} (e.g., thermal state). & A \textit{received, noise-corrupted coherent state} (original parameters and noise level unknown). \\
\hline
Primary Objective & High-fidelity generation of the \textit{predefined target state}. & High-fidelity restoration of an \textit{unknown input coherent state} from corruption. \\
\hline
\end{tabular}
\end{table*}

\subsection{Contributions}
\label{subsec:contributions}
Inspired by successful classical \cite{ho2020denoising,croitoru2023diffusion,zhang2024motiondiffuse}and discrete-variable (DV) quantum diffusion models \cite{cheng2024quantum, zhang2024QuDDPM, cacioppo2023quantumdm}, we introduce two novel models that leverage the continuous-variable (CV) quantum diffusion process to overcome critical limitations, such as operational rigidity, found in prior CV learning frameworks \cite{killoran2018continuous, shrivastava2019optical, cepaite2022born}. These are the \textbf{Continuous-Variable Quantum Diffusion Generative Model (CVQD-G)} for high-fidelity state generation, and the \textbf{Continuous-Variable Quantum Diffusion Restoration Model (CVQD-R)} for robust state recovery. The underlying step-wise, iterative refinement inherent to the quantum diffusion process enables both flexible state generation and particularly robust state restoration, marking a significant advancement for practical applications in noisy quantum systems.

While both CVQD-G and CVQD-R are built upon the same quantum diffusion process, they are tailored for distinct objectives. Their fundamental difference lies in their training data and operational starting points. CVQD-G is trained by diffusing a single, predefined target state and initiates generation from a standard noise state. In contrast, CVQD-R is trained on a diverse set of randomly sampled coherent states and begins its restoration process directly from a received, unknown, noise-corrupted state. These key distinctions are detailed in Table~\ref{tab:model_comparison_new_pos}.

The central idea of our framework, illustrated in Fig.~\ref{fig:main_idea}, is a learnable, bidirectional process driven by a physical noise model:
\begin{itemize}
    \item \textbf{Forward Diffusion:} A target quantum state $\rho_0$ is progressively degraded into a simple thermal state $\rho_T$ via repeated applications of a non-unitary \textbf{thermal loss channel}. This channel mimics natural decoherence (e.g., during optical fiber transmission) and is physically modeled as a beam splitter mixing the system qumode with the thermal environment.
    \item \textbf{Backward Restoration/Denoising:} The model learns to reverse this trajectory. Starting from $\rho_T$ (for generation) or a noisy state (for restoration), a CVQNN-based learnable operation progressively inverts the thermal loss at each step, reconstructing $\rho_0$ by incrementally removing noise.
\end{itemize}
This framework, by its very design of learning to reverse a defined noise process, inherently addresses both \textbf{state generation and restoration} within a single, coherent model. Our specific contributions are:
\begin{enumerate}

    \item \textbf{Novel CVQD-G Framework:} We develop a quantum generative model operating entirely in the CV domain. The forward process uses the physically driven thermal loss channel. The reverse process implements a learnable, non-unitary mapping via a CVQNN acting on the system and an ancillary qumode, followed by a trace-out operation. A key feature is a timestep embedding mechanism enabling parameter sharing across all diffusion steps, substantially reducing model complexity.

    \item \textbf{Application to State Recovery:} We demonstrate the framework's effectiveness in restoring target states from partially thermal mixed states. We also introduce the specialized variant, CVQD-R, trained specifically for restoration tasks. Our simulations show that CVQD-R can restore coherent states with unknown parameters from thermally degraded inputs, achieving fidelity exceeding 89\%, showcasing practical value for quantum communication.

    \item \textbf{Comprehensive Complexity Analysis:} We conduct a thorough analysis of the framework's computational costs, yielding detailed insights into its performance characteristics. For instance, this analysis reveals the training complexity to be $O(I \times B \times L)$ (where $I$ denotes iterations, $B$ represents batch size, and $L$ is the number of layers per step), which is notably independent of the total diffusion steps $T$ due to our efficient state transition method. This finding, in conjunction with favorable inference and space complexities, demonstrates the framework's efficiency and scalability, particularly when compared to methods like Optical GANs \cite{shrivastava2019optical} that incur additional measurement and discriminator overheads.

    \item \textbf{Verification of Robustness and Versatility:} Through numerical simulations, we verify the framework's effectiveness in generating various Gaussian (coherent, squeezed coherent) and non-Gaussian (Fock, cat) states with high fidelity (typically $>99\%$). We demonstrate its robustness by successfully generating target states even when starting the reverse process from partially noise-corrupted states across varying noise conditions ($\bar{n}=0$ and $\bar{n}=0.5$).
\end{enumerate}
In conclusion, this work introduces a flexible and powerful diffusion-based paradigm for CV quantum systems, offering unified solutions for high-fidelity state generation and robust state restoration in the presence of thermal noise.

The remainder of this article is structured as follows: Section~\ref{sec:Preliminary and related work} establishes  CV Quantum Computing  fundamentals and related quantum generative models. Section~\ref{sec:Continuous-variable Quantum Diffusion Model} presents a comprehensive description of our CVQD-G framework, including its underlying principles, quantum circuit design, and implementation methodology. Section~\ref{section:CVQD-R} presents our CVQD-G-based denoising methodology for quantum state restoration. Section~\ref{section:complex} analyzed the complexity of CVQD-G. Section~\ref{section:numerical_simulations} demonstrates numerical results for both state restoration and generation tasks. Section~\ref{section: conclusion} summarizes our contributions and discusses future directions. 

The notations used throughout the paper are summarized in Table~\ref{tab:notations}.

\begin{table}[t]
\centering
\caption{Notations.}
\label{tab:notations}
\begin{tabular}{|c|l|}
\hline
\textbf{Notation} & \textbf{Description} \\ \hline
$t$               & Timestep \\ \hline
$T$               & Total timesteps of the Diffusion Process \\ \hline
$|\psi\rangle$    & Quantum State \\ \hline
$\rho$            & Density Matrix of a Quantum System \\ \hline
$\rho_0$          & Target Quantum State \\ \hline
$\rho_t$          & Quantum State at Timestep $t$ \\ \hline
$\rho_T$          & Quantum State at the End of Diffusion Process \\ \hline
$\mathcal{T}(t)$  & Time Embedding Circuit at Timestep $t$ \\ \hline
$f_\vartheta(\rho_t, t)$& Denoising Process Function Parameterized by $\vartheta$\\ \hline
$\eta_t$          & Transmissivity at Timestep $t$ \\ \hline
$\bar{\eta}_t$    & Accumulated Noise Parameter $\bar{\eta}_t = \prod_{i=1}^t \eta_i$ \\ \hline
$\mathcal{E}(\cdot)$ & Diffusion Process Function \\ \hline
$U(\vartheta)$& Denoising Circuit Parameterized by $\vartheta$\\ \hline
$I$               & Identical Matrix \\ \hline
$\text{tr}(\cdot)$ & Trace Operation \\ \hline
$F(\cdot, \cdot)$  & Quantum Fidelity Function \\ \hline
$\lambda$         & Hyperparameter in Loss Function \\ \hline
$\mathcal{L}_t$   & Loss Function at Timestep $t$ \\ \hline
$\mathcal{L}_{\text{total}}$ & Total Loss Function \\ \hline
$\tau_t$          & Time Embedding Quantum State \\ \hline
\end{tabular}
\end{table}

\section{Preliminary and Related Works}
\label{sec:Preliminary and related work}

\subsection{Preliminary}

This section briefly introduces four basic concepts  in continuous-variable quantum computing: qumode representations as the fundamental information carriers, Gaussian states and operations as the practical building blocks for CV quantum computing, and the thermal loss channel as the primary noise model affecting quantum information processing in realistic environments.

\subsubsection{Qumode Representations}
 CV Quantum Computing utilizes continuous physical observables, such as position \( x \), contrasting with discrete-variable quantum computing. In  CV Quantum Computing, quantum information is encoded in optical modes called \emph{qumodes}, which serve as the fundamental units of quantum circuits.

A qumode state can be represented through two complementary frameworks: Fock space and phase space\cite{braunstein2005quantum}.

Fock space representation provides a direct description of the energy distribution in the optical field. In this representation, a qumode state is described by its energy distribution using a density matrix in the infinite basis \( \{ |n\rangle \}_{n=0}^\infty \), where \( |n\rangle \) denotes a state with \( n \) photons.

Phase space representation provides an intuitive alternative using position \( x \) and momentum \( p \) as conjugate coordinates, mapping a qumode state to \( \mathbb{R}^2 \). The quantum state is characterized by the \emph{Wigner function}\cite{wigner1932wigner}:

\begin{equation}
W(x, p) = \frac{1}{\pi \hbar} \int_{-\infty}^{\infty} \langle x + y | \rho | x - y \rangle e^{-2ipy / \hbar} \, dy,
\end{equation}
where \( \rho \) is the density matrix, \(\hbar\) is the reduced Planck constant, and \( y \) is the integration variable. This function completely describes quantum states in phase space.

\subsubsection{Gaussian States:}

Gaussian states are quantum states whose Wigner functions exhibit Gaussian distributions in phase space\cite{WANG2007Gaussian}. They serve as building blocks for more complex quantum states in continuous-variable quantum computing. Several important Gaussian states are described below:

\textbf{Vacuum State:}
The vacuum state represents the ground state of the quantum harmonic oscillator, corresponding to the absence of photons in a quantum optical mode. It is the lowest energy state of a qumode and serves as the starting point for many quantum operations. In phase space, the vacuum state appears as a minimum uncertainty Gaussian distribution centered at the origin.

\textbf{Thermal State:} 
Thermal states represent quantum systems in thermal equilibrium at a finite temperature. They describe the natural state of environmental noise in quantum optical systems. Physically, thermal states consist of a statistical mixture of photon number states, with occupation probabilities following a Boltzmann distribution. In phase space, thermal states appear as two-dimensional Gaussian distributions with variance \(\left(\bar{n} + \frac{1}{2}\right)I_2\), where \(\bar{n}\) represents the average photon number. The thermal state can be formally expressed as:
\begin{equation}
\rho_{\text{th}} = \rho_G\left( 0, \left(\bar{n} + \frac{1}{2}\right)I_2 \right),
\end{equation}
where \(\rho_G(\cdot, \cdot)\) represents a Gaussian state with zero mean and a covariance matrix of \(\left(\bar{n} + \frac{1}{2}\right)I_2\), with \(I_2\) as the two-dimensional identity matrix. When \(\bar{n} = 0\), the thermal state reduces to the vacuum state, \(\rho_{\text{th}} = |0\rangle\langle 0|\). When \(\bar{n} \neq 0\), the thermal state becomes a mixed state, reflecting entropy and randomness introduced by thermal noise.

\textbf{Coherent States:}
Coherent states $|\alpha\rangle$ are the quantum mechanical analogs of classical electromagnetic waves. They are eigenstates of the annihilation operator $\hat{a}$:
\begin{equation}
\hat{a}|\alpha\rangle = \alpha|\alpha\rangle.
\end{equation}
The complex parameter $\alpha$ represents the displacement from the origin in phase space. Coherent states maintain the minimum uncertainty product allowed by Heisenberg's uncertainty principle, with equal uncertainties in both quadratures. In Fock space, a coherent state can be expressed as:
\begin{equation}
|\alpha\rangle = e^{-|\alpha|^2/2} \sum_{n=0}^\infty \frac{\alpha^n}{\sqrt{n!}} |n\rangle.
\end{equation}

\textbf{Squeezed States:}
Squeezed states represent non-classical light with asymmetric quantum noise distribution. They exhibit reduced quantum noise in one quadrature (position or momentum) at the expense of increased noise in the conjugate quadrature, enabling measurements with precision beyond the standard quantum limit.

\subsubsection{Gaussian Operators}

Gaussian operations are transformations that map a Gaussian state to another Gaussian state. In quantum optical implementations, these operators correspond to physical processes that can be realized using linear optical components, parametric processes, and homodyne measurements. Below are the fundamental Gaussian operators that form the building blocks for CV quantum computing:

\textbf{Displacement Operator:}
The displacement operator shifts a quantum state in phase space without changing its shape or uncertainty properties.  For a complex parameter \(\alpha \in \mathbb{C}\), the displacement operator transforms the phase space coordinates as:
\begin{equation}
D(\alpha): 
\begin{bmatrix}
x \\
p
\end{bmatrix}
\mapsto 
\begin{bmatrix}
x + \sqrt{2}\operatorname{Re}(\alpha) \\
p + \sqrt{2}\operatorname{Im}(\alpha)
\end{bmatrix}.
\end{equation}
When applied to the vacuum state, the displacement operator generates coherent states: \(D(\alpha)|0\rangle = |\alpha\rangle\).

\textbf{Phase Shift Operator:}
The phase shift operator rotates a quantum state around the origin of phase space. For a phase angle \(\phi\in[0,2\pi]\), the transformation is:
\begin{equation}
R(\phi): 
\begin{bmatrix}
x \\
p
\end{bmatrix}
\mapsto 
\begin{bmatrix}
\cos\phi & \sin\phi \\
-\sin\phi & \cos\phi
\end{bmatrix}
\begin{bmatrix}
x \\
p
\end{bmatrix}.
\end{equation}

\textbf{Squeezing Operator:}
The squeezing operator reshapes the uncertainty distribution of a quantum state, reducing noise in one quadrature while increasing it in the conjugate quadrature. For a squeezing parameter \( r \in \mathbb{R} \), the transformation is:
\begin{equation}
S(r): 
\begin{bmatrix}
x \\
p
\end{bmatrix}
\mapsto 
\begin{bmatrix}
e^{-r} & 0 \\
0 & e^{r}
\end{bmatrix}
\begin{bmatrix}
x \\
p
\end{bmatrix}.
\end{equation}

\textbf{Beam Splitter Operator:}
The beam splitter operator acts on two modes simultaneously, creating linear combinations of their quadratures. It transforms the input modes into weighted superpositions, where each output mode contains components from both input modes. For two qumodes \(\hat{a}, \hat{b}\) and a transmissivity parameter \(\eta\), the transformation is:
\begin{equation}
U_{\text{BS}}(\eta): \begin{pmatrix} \hat{a} \\ \hat{b} \end{pmatrix} \mapsto \begin{pmatrix} \sqrt{\eta}\hat{a} + \sqrt{1-\eta}\hat{b} \\ -\sqrt{1-\eta}\hat{a} + \sqrt{\eta}\hat{b} \end{pmatrix}.
\end{equation}

\subsection{Related Works}

We organize our discussion of related work along two key research directions: continuous-variable quantum generative models and quantum adaptations of diffusion models.

\textbf{Continuous-Variable Quantum Generative Models.} Continuous-Variable Quantum Generative Models (CVQGMs) represent a significant branch of quantum machine learning, leveraging the infinite-dimensional Hilbert space of CV systems to efficiently represent continuous distributions. Killoran et al. \cite{killoran2018continuous} pioneered the Continuous-Variable Quantum Neural Network (CVQNN) framework, employing parameterized quantum optical circuits that have since been extended to quantum state preparation \cite{Arrazola_2019}, machine learning \cite{Killoran2019strawberryfields}, and cryptography \cite{shi2020cryptographyCVQNN}.

Several specialized architectures have emerged within this paradigm. The Optical Generative Adversarial Network (OpticalGAN) \cite{shrivastava2019optical} adapts the GAN framework to photonic hardware, utilizing optical parametric processes for adversarial training. The Continuous-Variable Born Machine (CVBM) \cite{cepaite2022born} efficiently represents and learns both quantum and classical continuous distributions through the CV quantum computing architecture. Similarly, the Continuous-Variable Quantum Boltzmann Machine (CVQBM) \cite{bangar2024boltzmann} employs energy-based neural networks within CV photonic quantum computers, using quantum imaginary time evolution to prepare thermal states for generating continuous probability distributions.

While these models have demonstrated impressive capabilities in generating complex quantum states, they have primarily focused on state generation tasks. The equally important challenge of restoring quantum states that have been corrupted by thermal noise has received less attention. Existing CV quantum generative models \cite{shrivastava2019optical, cepaite2022born, braunstein2005quantum} typically excel at generation but lack demonstrated restoration capabilities. Our CVQD-G uniquely addresses both capabilities by employing a thermal loss channel for forward diffusion coupled with a CVQNN-based reverse denoising circuit.

\textbf{Quantum Adaptations of Diffusion Models.} Classical diffusion models, inspired by thermal equilibrium processes, have been remarkably successful in generative tasks. Several research groups have recently developed quantum analogs across different paradigms. Zhang et al. \cite{zhang2024QuDDPM} introduced QuDDPM (Quantum Denoising Diffusion Probabilistic Models), focusing on generating individual pure states from target quantum distributions. In a parallel development, Chen et al. \cite{cheng2024quantum} pioneered a quantum diffusion framework for discrete-variable systems that utilizes depolarizing channels as the quantum noise mechanism, demonstrating the ability to generate both pure and mixed quantum states.

However, these approaches predominantly operate in qubit-based systems or implement hybrid frameworks that combine classical diffusion processes with quantum denoising components \cite{cheng2024quantum, zhang2024QuDDPM, cacioppo2023quantumdm}. Our work addresses this gap by developing a diffusion model that operates entirely within the continuous-variable quantum framework, offering both generation and restoration capabilities not demonstrated in previous CV quantum generative models.

\section{Continuous-variable Quantum Diffusion Model For State Generation}
\label{sec:Continuous-variable Quantum Diffusion Model}

\begin{figure*}[t]
	\centering
	\includegraphics[width=\linewidth]{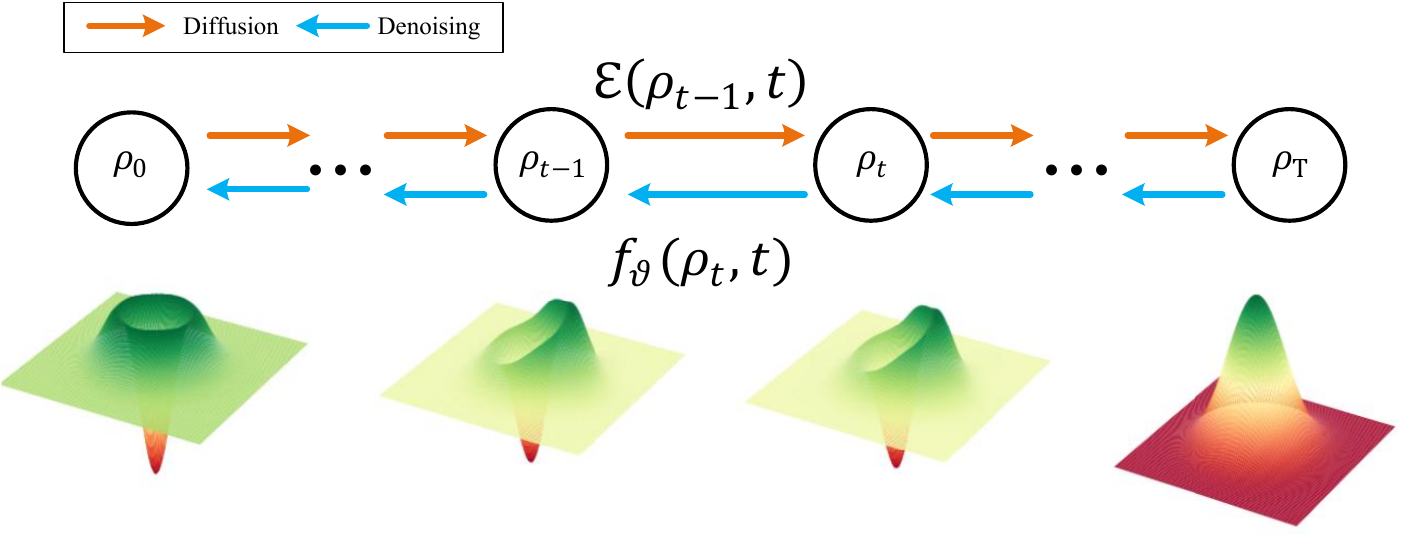}
	\caption{The framework of the proposed Continuous-Variable Quantum Diffusion Generative Model (CVQD-G).}
	\label{fig:general_framework}
\end{figure*}

\subsection{General Framework}
Fig.~\hyperref[fig:general_framework]{2} shows a general framework of the Continuous-Variable Quantum Diffusion Generative Model (CVQD-G). As in classical diffusion models, CVQD-G has both a forward (diffusion) process $\mathcal{E}(\rho_{t-1}, t)$ and a backward (denoising) process $f_\vartheta(\rho_t, t)$.

In the forward diffusion process, environmental thermal noise is progressively added to the original state $\rho_0$ through multiple timesteps $t$. This process is modeled by a thermal loss channel (visualized as a virtual beam splitter in Fig. \ref{fig:main_idea}) with the input state $\rho_{t-1}$ and a loss parameter $\eta_t$ . When $t=0$, $\rho_0$ represents the target state, while at $t=T$, $\rho_T$ represents the thermal state dependent on the environment's mean photon number $\bar{n}$.

The backward denoising process employs a model $f_\vartheta(\rho_t, t)$ with trainable parameters $\vartheta$ and time embedding. Given a noisy quantum state $\rho_t$ at timestep $t$, this trainable circuit generates a quantum state $\rho_{t-1}$ that is closer to the original state $\rho_0$. During training, the model parameters $\vartheta$ are optimized by comparing the predicted state with the actual $\rho_{t-1}$ obtained from the forward process. By sequentially applying the trained function $f_\vartheta(\cdot, \cdot)$ $T$ times to $\rho_T$, the target quantum state is progressively recovered.

\subsection{Forward Diffusion Process}

Diffusion models draw inspiration from fundamental physics principles in thermodynamic where systems naturally evolve from ordered states toward thermal equilibrium states over time \cite{groot2013thermodynamics}. CVQD-G applies this concept to quantum systems by employing thermal loss channels to gradually transform pure quantum states into thermal equilibrium states. This approach effectively models environmental noise effects commonly encountered in quantum communication \cite{WANG2007Gaussian}, providing both theoretical elegance and practical relevance to real-world quantum systems.

The thermal loss channel can be intuitively understood in the Heisenberg picture as a beam splitter interaction between the quantum system and a thermal environment  (as Fig.\ref{fig:main_idea})\cite{gisin2007quantumcomm} . When an optical mode $\hat{a}_{t-1}$ interacts with an environmental mode $\hat{e}_t$ prepared in a thermal state, the transformation at timestep $t$ is given by:
\begin{equation}
\hat{a}_t = \sqrt{\eta_t} \, \hat{a}_{t-1} + \sqrt{1 - \eta_t} \, \hat{e}_t,
\label{equ:diffusion Heisenberg}
\end{equation}
where $\eta_t \in [0, 1]$ is the timestep-dependent transmissivity parameter. This formulation clearly illustrates how the output mode becomes a weighted combination of the input quantum state and thermal noise.

The transmissivity parameter $\eta_t$ controls the strength of the interaction with the thermal environment: when $\eta_t = 1$, the quantum state remains unchanged, while at $\eta_t = 0$, the output becomes entirely a thermal state. For intermediate values, the channel produces a partial mixing of the input state with thermal noise. This non-unitary process increases the system's entropy \cite{Holevo_2012}, gradually transforming a potentially low-entropy state (e.g., a pure state) into a higher-entropy state (e.g., a mixed state) as energy and coherence are lost to the environment.

Equivalently, the forward diffusion process can be expressed in the Schrödinger picture using density operators. At each timestep $t$, the thermal loss channel transforms the quantum state $\rho_{t-1}$ as follows:
\begin{equation}
\rho_t = \mathcal{E}(\rho_{t-1},\eta_t) = \mathrm{Tr}_{\text{E}}[U_{\text{BS}}(\eta_t)(\rho_{t-1} \otimes \rho_{\text{th}}(\bar{n}))U_{\text{BS}}^{\dagger}(\eta_t)]
\label{equ:diffusion Schrödinger}
\end{equation}
where $\rho_{t-1}$ is the input quantum state, $\rho_{\text{th}}(\bar{n})$ denotes the environmental thermal state with mean photon number $\bar{n}$, and $U_{\text{BS}}(\eta_t)$ is the beam splitter operator. The operation $\mathrm{Tr}_{\text{E}}$ indicates tracing out the environmental degrees of freedom, capturing the non-unitary nature of this open quantum system evolution.

By designing an appropriate sequence $\{\eta_t\}$, the thermal loss channel progressively transforms the original state $\rho_0$ toward the thermal state, eventually yielding $\rho_T \approx \rho_{\text{th}}(\bar{n})$ when $T$ is sufficiently large. This controlled diffusion toward thermal equilibrium due to information loss when coupling with the thermal environment forms the basis of the forward diffusion process in CVQD-G.

\subsubsection{Efficient Quantum State Transition to Arbitrary Timestep}

In the diffusion process of our model, transitioning from the initial state $\rho_0$ to any arbitrary timestep state $\rho_t$ conventionally requires iteratively applying the thermal loss channel $\mathcal{E}(\rho_{t-1},\eta_t)$ a total of $t$ times, following Eqs.~\ref{equ:diffusion Heisenberg} and ~\ref{equ:diffusion Schrödinger}. This sequential approach becomes computationally prohibitive as $t$ increases, creating a significant bottleneck during both training and sampling phases.

To overcome this limitation, we develop a direct transformation method in CVQD-G that enables single-step computation of any timestep state $\rho_t$ directly from $\rho_0$. This advancement can be formulated using two complementary mathematical formalisms. The Heisenberg picture, with its focus on operator transformations, provides an elegant framework for proving the theoretical validity of our approach. Meanwhile, the Schrödinger picture offers a direct characterization of density matrix evolution, making it particularly suitable for numerical implementation within our quantum framework.

\textbf{Theorem 1:} In CVQD-G, the system state at timestep $t$ (denoted $\hat{a}_t$ in the Heisenberg picture or $\rho_t$ in the Schrödinger picture) can be obtained directly from the corresponding initial system state ($\hat{a}_0$ or $\rho_0$) using a single effective thermal loss channel with appropriately modified parameters:

\medskip 

 1) In the Heisenberg picture, describing the system annihilation operator:
    \begin{equation}
    \hat{a}_t = \sqrt{\bar{\eta}_t} \, \hat{a}_0 + \sqrt{1 - \bar{\eta}_t} \, \hat{E}_t,
    \label{eq:theorem_heisenberg_revised}
    \end{equation}
    where $\bar{\eta}_t = \prod_{i=1}^{t} \eta_i$ is the effective cumulative transmissivity (each $\eta_i$ being the transmissivity of an individual diffusion step). $\hat{E}_t$ is an effective thermal environmental mode operator, reflecting interaction with an environment of constant mean photon number $\bar{n}$ (common to all $t$ steps).

\medskip

 2) In the Schrödinger picture, describing the system density matrix:
    \begin{equation}
    \rho_t = \mathrm{Tr}_{\text{E}}[U_{\text{BS}}(\bar{\eta}_t)(\rho_{0} \otimes \rho_{\text{th}}(\bar{n}))U_{\text{BS}}^{\dagger}(\bar{\eta}_t)],
    \label{eq:theorem_schrodinger_revised}
    \end{equation}
    where $U_{\text{BS}}(\bar{\eta}_t)$ denotes the beam splitter unitary with transmissivity $\bar{\eta}_t$, $\rho_{\text{th}}(\bar{n})$ is the thermal state of the ancillary environmental mode (with said mean photon number $\bar{n}$), and $\mathrm{Tr}_{\text{E}}$ is the partial trace over this environment.
\medskip

\noindent\textbf{Proof:}
The proof first establishes, via induction in the Heisenberg picture, that the $t$-step evolution of the system's annihilation operator simplifies to a single interaction with an effective thermal environment. This Heisenberg result is then mapped to the Schrödinger picture, confirming that the $t$-step density matrix is equivalent to applying a single effective thermal loss channel to the initial state.

The complete proof is provided in Appendix~\ref{app:theorem1_proof}.\hfill$\square$

This direct transformation significantly enhances computational efficiency by eliminating the need to sequentially compute all intermediate states. During training and sampling, we can now instantly access quantum states at arbitrary timesteps, which proves particularly valuable when working with large diffusion lengths. The formulation also provides theoretical insights into the cumulative effects of thermal noise on quantum states, supporting both our analytical understanding and practical implementation of the model.

\subsubsection{Noise Schedule}
The rate of diffusion towards the thermal state \( \rho_T \) is governed by the noise schedule, defined by the single-step transmissivities \( \{\eta_t\}_{t=1}^T \). Each \( \eta_t \) must be in \( [0, 1] \), and the schedule ensures the cumulative transmissivity \( \bar{\eta}_T = \prod_{i=1}^{T} \eta_i \) becomes small, indicating significant diffusion.

In this work, we employ a linear schedule for \( \eta_t \):
\begin{equation}\label{eq:linear_schedule} 
\eta_t = \eta_{0} + (\eta_{T} - \eta_{0}) \frac{t}{T}.
\end{equation}
Here, \( \eta_0 \) and \( \eta_T \) are hyperparameters that define the nominal start and end points for this linear interpolation across the \( T \) steps. Their values are chosen to ensure \( \eta_t \) remains physically valid and \( \bar{\eta}_T \approx 0 \).

\subsection{Backward Denoising Process}
The backward process in CVQD-G aims to systematically recover the original quantum state $\rho_0$ from the environment's thermal state $\rho_T$. This recovery process faces two fundamental theoretical challenges:

The first challenge concerns parametric efficiency: While the forward diffusion process utilizes the thermal loss channel to directly compute any $\rho_t$ using known parameters $\bar{n}$ and $\bar{\eta}_t$, the backward process is fundamentally a training procedure. Unlike the forward process, we cannot employ a quantum channel model to directly obtain states at arbitrary timesteps. This would naively require training $T$ timestep-specific denoising functions $\{f_{\vartheta_t}(\rho_t)\}_{t=1}^T$, imposing excessive computational demands.

The second challenge arises from the inherently non-unitary forward diffusion process: as quantum information irreversibly dissipates into the environment, the backward process must employ non-unitary operations \cite{kraus1971general}.

\begin{figure*}[ht]
\centering
\includegraphics[width=\linewidth]{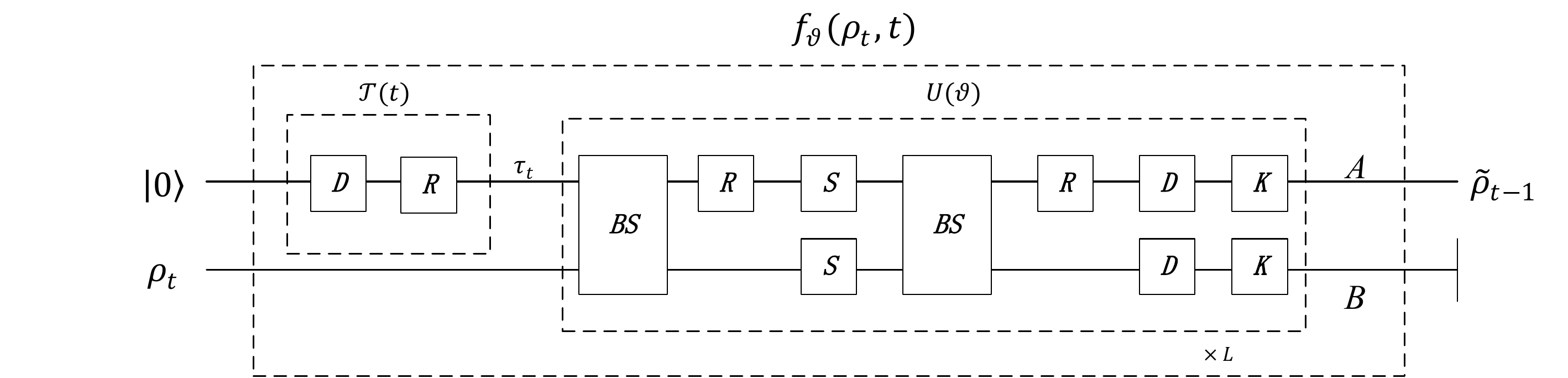}
\caption{Schematic of the CVQD-G denoising framework. The function $f_\vartheta(\rho_t, t)$ consists of a time embedding circuit $\mathcal{T}(t)$ and a denoising circuit $U(\vartheta)$.}
\label{fig:architecture}
\end{figure*}
    
Our solution, illustrated in Figure \ref{fig:architecture}, addresses both challenges through a unified architecture $f_\vartheta(\rho_t, t)$. This architecture consists of a time embedding module $\mathcal{T}(t)$ and a parameterized denoising circuit $U(\vartheta)$. For parametric efficiency, we implement timestep embedding $\mathcal{T}(t)$ that generates a timestep-dependent state $\tau_t$ by making $t$ an explicit input to our model, transforming the denoising approach from a set with $T$ groups of separate parameters $\{f_{\vartheta_t}(\rho_t)\}_{t=1}^T$ to a single shared parameter set $f_\vartheta(\rho_t, t)$ conditioned on timestep. To enable non-unitary operations, we introduce ancillary qumodes that extend beyond the primary quantum system. The denoising process works by preparing the composite system $\tau_t \otimes \rho_t$, applying $U(\vartheta)$, and achieves non-unitary transformation by obtaining the predicted state $\tilde{\rho}_{t-1}$ through tracing out the auxiliary qumode $B$ while retaining qumode $A$.

\subsubsection{Time Embedding Circuit}

To address the first challenge of parametric efficiency, we developed a time embedding mechanism that leverages the natural phase evolution of coherent states. This approach allows us to condition a single model on different timesteps rather than training separate models for each timestep. The coherent states evolve according to:

\begin{equation}
\alpha(t) = e^{-i\omega t}\alpha.
\end{equation}

For any non-zero coherent state $|\alpha\rangle$, this phase evolution provides an elegant way to encode timestep information. As shown in the $\mathcal{T}(t)$ component of Fig.~\ref{fig:architecture}, we implement this through two operations: first, a displacement gate $D(\alpha)$ with fixed parameter $\alpha \in \mathbb{R}^{+}$ shifts the vacuum state along the x-axis in phase space; then, a rotation gate \(R(\phi(t))\) applies a timestep-dependent phase $\phi(t) = \frac{t\pi}{T}$, linearly mapping discrete timestep $t$ to phase angle $\phi \in [0,\pi]$. This approach effectively embeds the timestep information directly into the quantum state itself.

\subsubsection{Denoising Circuit}

To overcome the second challenge of implementing non-unitary operations, we designed a denoising circuit that operates on an extended Hilbert space. The denoising circuit $U(\vartheta)$ acts on the composite system $\tau_t \otimes \rho_t$, a two-qumode system. After applying this circuit, the predicted output $\widetilde{\rho}_{t-1}$ is obtained by tracing out subsystem $B$:

\begin{equation}
\widetilde\rho_{t-1} = f_\vartheta(\rho_t, t) = \text{tr}_B\left(U(\vartheta)(\tau_t \otimes \rho_t) U^\dagger(\vartheta)\right).
\end{equation}

While using the same qumode for both input $\rho_t$ and output $\widetilde{\rho}_{t-1}$ might seem intuitive, this approach can be problematic. Since the difference between states $\rho_t$ and $\rho_{t-1}$ may be minimal, the circuit could default to simply copying its input rather than generating the appropriate denoised output \cite{cheng2024quantum}.

As shown in the $U(\vartheta)$ component of Fig.~\ref{fig:architecture}, our denoising circuit consists of $L$ layers of two-qumode CVQNN circuits. Each CVQNN layer comprises two distinct parts \cite{killoran2018continuous}, analogous to how classical neural networks require both linear and nonlinear components:
\begin{itemize}
    \item A linear component utilizing Gaussian operators, including Displacement gate \(D\), Rotation gate \(R\), Squeeze gate \(S\), and Beam splitter gate \(BS\)
    \item A nonlinear component implemented through the Kerr gate \(K\), defined as:
    \begin{equation}
    K(\kappa) = \exp\left( i\kappa (\hat{a}^\dagger \hat{a})^2 \right),
    \end{equation}
    where $\kappa$ is the gate parameter.
\end{itemize}

Our architecture leverages the fundamental result that combining linear operations with a single nonlinear operation enables universal CV quantum computation with polynomial overhead \cite{lloyd1999quantum, Arrazola_2019}. Gaussian operators alone are restricted to generating only quadratic Hamiltonians, limiting their computational expressivity. The Kerr gate provides the necessary nonlinearity to construct higher-order Hamiltonians, enabling universal quantum computation beyond Gaussian limitations. Additionally, being diagonal in the Fock basis, the Kerr gate facilitates faster and more reliable numerical simulations \cite{Arrazola_2019}.

\subsection{Training and Generation}
The fundamental goal of training CVQD-G is to optimize the denoising circuit parameters $\vartheta$ such that the predicted state $\tilde{\rho}_{t-1}$ closely approximates the actual state $\rho_{t-1}$. This is achieved by maximizing the quantum fidelity between these states.

Our total loss function is defined as:
\begin{equation}
\mathcal{L}_{\text{total}} = \mathcal{L}_0 + \lambda \frac{1}{\mathbb{B}} \sum_{i=1}^{\mathbb{B}} \mathcal{L}_{t-1},
\end{equation}
where $\mathbb{B}$ represents the batch size and $\lambda$ is a hyperparameter balancing the different loss components. The primary component $\mathcal{L}_{t-1}$ is computed as:
\begin{equation}
\mathcal{L}_{t-1} = 1 - F(\rho_{t-1}, \tilde{\rho}_{t-1}) + \gamma P(\tilde{\rho}_{t-1}).
\end{equation}

Here, $F(\cdot, \cdot)$ is the quantum fidelity function, which measures similarity between quantum states:
\begin{equation}
F(\rho, \sigma) = \left( \mathrm{tr} \sqrt{\sqrt{\rho} \sigma \sqrt{\rho}} \right)^2.
\end{equation}

For numerical implementation, we use the Fock backend in \textbf{Strawberry Fields} \cite{Killoran2019strawberryfields}, which necessarily truncates the infinite-dimensional Hilbert space. This truncation can lead to issues where circuit operations push states beyond the computational basis, resulting in unnormalized density matrices. To address this challenge, we incorporate a penalty term $P(\tilde{\rho}_{t-1})$ in the loss function:
\begin{equation}
P(\tilde{\rho}_{t-1}) = \left(\text{tr}(\tilde{\rho}_{t-1}) - 1 \right)^2,
\end{equation}
which ensures the trace of the output density matrix remains close to 1.

The complete CVQD-G training procedure is outlined in \textbf{Algorithm 1}. During each iteration, we calculate $\mathcal{L}_{\text{total}}$ and update the parameters $\vartheta$ using gradient-based optimization until convergence or reaching a maximum iteration count.

Once the model is trained, we can leverage the optimized parameters $\vartheta$ to perform quantum state generation. Starting from a thermal state $\rho_T$, the backward process sequentially applies the denoising circuit to recover the initial state $\rho_0$, as detailed in \textbf{Algorithm 2}.

\subsection{Algorithms}
The CVQD-G training algorithm is presented in Algorithm 1. In each iteration, \( \mathcal{L}_{\text{total}} \) is calculated to update the trainable parameters \( \vartheta \) until \( \mathcal{L}_{\text{total}} \) converges or the maximum number of iterations is reached.

\begin{algorithm}[h]
\caption{CVQD-G Forward Diffusion Training Algorithm}
\textbf{Input:} Transmissivity schedule parameters \(\{\bar{\eta}_t\}_{t=1}^T\), total timestep \(T\), initial state \(\rho_0\), average photon number of thermal state \(\bar{n}\), batch size \(\mathbb{B}\), maximum iteration number, learning rate \(\iota\), weight parameter \(\lambda\). \\

\textbf{1.} \(\vartheta\) starts with a small number near 0. \\
\textbf{2.} \(\rho_{\text{th}} \gets \rho_G(0, (\bar{n} + \frac{1}{2})I_2)\). \\
\textbf{3.} \textbf{repeat} \\
\textbf{4.} \(\rho_1 \gets \text{tr}_E(U_{\text{BS}}(\bar{\eta}_1)(\rho_0 \otimes \rho_{\text{th}}(\bar{n}))U_{\text{BS}}^\dagger(\bar{\eta}_1))\).\\
\textbf{5.} \(\tau_1 \gets \mathcal{T}(1) |0\rangle \langle 0| \mathcal{T}^\dagger(1)\). \\
\textbf{6.} \(\tilde{\rho}_0 \gets \text{tr}_B(U(\vartheta)(\tau_1 \otimes \rho_1)U^\dagger(\vartheta))\). \\
\textbf{7.} \(\mathcal{L}_0 \gets 1 - F(\rho_0, \tilde{\rho}_0) + \gamma P(\tilde{\rho}_0)\). \\
\textbf{8.} \(\mathcal{L}_{\text{total}} \gets 0\). \\
\textbf{9.} \textbf{for} \(i \text{ from } 1 \text{ to } \mathbb{B}\) \textbf{do} \\
\textbf{10.} \quad Sample a unique timestep \(t\) from the uniform distribution \(\{2,3,\ldots,T\}\). \\
\textbf{11.} \quad \(\rho_t \gets \text{tr}_E(U_{\text{BS}}(\bar{\eta}_t)(\rho_0 \otimes \rho_{\text{th}}(\bar{n}))U_{\text{BS}}^\dagger(\bar{\eta}_t))\).\\
\textbf{12.} \quad \(\tau_t \gets \mathcal{T}(t) |0\rangle \langle 0| \mathcal{T}^\dagger(t)\). \\
\textbf{13.} \quad \(\tilde{\rho}_t \gets \text{tr}_B(U(\vartheta)(\tau_t \otimes \rho_t)U^\dagger(\vartheta))\). \\
\textbf{14.} \quad \(\mathcal{L}_{\text{total}} \gets \mathcal{L}_{\text{total}} + 1 - F(\rho_{t-1}, \tilde{\rho}_{t-1}) + \gamma P(\tilde{\rho}_{t-1})\). 

\textbf{15.} \textbf{end for} \\
\textbf{16.} \(\mathcal{L}_{\text{total}} \gets \mathcal{L}_0 + \lambda \frac{1}{\mathbb{B}} \mathcal{L}_{\text{total}}\). \\
\textbf{17.} \(\vartheta\gets \vartheta - \iota \nabla_\vartheta \mathcal{L}_{\text{total}}\). \\
\textbf{18.} \textbf{until} \(\mathcal{L}_{\text{total}}\) converges or maximum number of iterations is reached. \\
\textbf{Output:} Optimal parameter \(\vartheta\).
\end{algorithm}

\begin{algorithm}[h]
\caption{CVQD-G Backward Denoising Generation Algorithm}
\textbf{Input:} Total timestep \(T\). Average photon number of thermal state \(\bar{n}\). Optimal parameter \(\vartheta\). \\

\textbf{1.} Initialize \(\tilde{\rho}\) as the thermal state: \\
\(\tilde{\rho} \gets \rho_{\text{th}} = \rho_G(0, (\bar{n} + \frac{1}{2})I_2)\). \\
\textbf{2.} \textbf{for} \(t\) from \(T\) to \(1\) \textbf{do} \\
\textbf{3.} \quad \(\tau_t \gets \mathcal{T}(t) |0\rangle \langle 0| \mathcal{T}^\dagger(t)\). \\
\textbf{4.} \quad Update the generation state: \\
\quad \(\tilde{\rho} \gets \text{tr}_B(U(\vartheta)(\tau_t \otimes \tilde{\rho})U^\dagger(\vartheta))\) using the previous value of \(\tilde{\rho}\). \\
\textbf{5.} \textbf{end for} \\
\textbf{Output:} The final generation state \(\tilde{\rho}_0 = \tilde{\rho}\).
\end{algorithm}

\section{Continuous-Variable Quantum Diffusion Model For State Restoration} 
\label{section:CVQD-R}

While the CVQD-G model is optimized for generating specific, predefined states, the more general challenge of restoring states with unknown original parameters from unknown levels of noise is addressed by the \textbf{Continuous-Variable Quantum Diffusion Restoration Model (CVQD-R)}. To achieve this, CVQD-R employs a distinct training strategy: instead of learning to reproduce a single target, it is trained on a diverse ensemble of randomly sampled coherent states, equipping it with the generalization capability needed to reverse thermal degradation without \textit{a priori} knowledge of the specific initial state or the precise noise characteristics.

\subsection{The Challenge of State Restoration in Noisy Channels}

In practical quantum communication, states transmitted by a sender (Alice) are inevitably corrupted by environmental noise before reaching the receiver (Bob)\cite{gisin2007quantumcomm, Holevo_2012}. 
Consider Alice sending a coherent state $\rho_{\text{in}}$ (e.g., $|\alpha_{\text{in}}\rangle\langle\alpha_{\text{in}}|$). Its passage through a noisy quantum channel, often modeled as a \textbf{thermal loss channel}, degrades it. 
This channel, characterized by transmissivity $\eta_{\text{ch}}$ and environmental mean photon number $\bar{n}_{\text{env}}$, transforms $\rho_{\text{in}}$ into a corrupted state $\rho_{\text{out}}$: 
\begin{equation}
\rho_{\text{out}} =\mathcal{E}(\rho_{out},\eta_{ch})= \mathrm{Tr}_{\text{E}}[U_{\text{BS}}(\eta_{\text{ch}})(\rho_{\text{in}} \otimes \rho_{\text{th}}(\bar{n}_{\text{env}}))U_{\text{BS}}^{\dagger}(\eta_{\text{ch}})],
\label{eq:thermal_channel_corruption}
\end{equation}
where $U_{\text{BS}}$ is the beam splitter interaction with the environmental thermal state $\rho_{\text{th}}(\bar{n}_{\text{env}})$, and $\mathrm{Tr}_{\text{E}}$ denotes the partial trace over the environment. 
Bob's task is to restore $\rho_{\text{out}}$ to retrieve Alice's original information. 
This is challenging because Bob's knowledge of Alice's state parameters or the precise noise characteristics is often incomplete. 
Our work focuses on effectively processing $\rho_{\text{out}}$ to recover the underlying quantum information under these practical constraints. 

\subsection{Theoretical Basis for Restoration}

The CVQD-R framework trains a quantum neural network to recognize and reverse the systematic effects of noise on quantum states. 
By learning the relationship between state properties and noise parameters, the model can infer the extent of corruption and apply a learned restoration operation\cite{Arrazola_2019, ho2020denoising}. 

Understanding how the variance of a quantum state is affected when traversing a thermal loss channel is crucial. When a coherent state passes through such a channel, characterized by transmissivity $\eta$ and an environment with mean photon number $\bar{n} > 0$, its output quadrature variance, $\langle (\Delta \hat{X})^2 \rangle_{\text{total}}$, is fundamentally a weighted combination of the initial coherent state's intrinsic variance and the variance contributed by the thermal environment. This governing relationship is given by\cite{braunstein2005quantum, WANG2007Gaussian}:
\begin{equation}
\langle (\Delta \hat{X})^2 \rangle_{\text{total}} = \eta \langle (\Delta \hat{X}_{\text{coherent}})^2 \rangle + (1-\eta) \langle (\Delta \hat{X}_{\text{thermal}})^2 \rangle.
\label{eq:total_variance_combination} 
\end{equation}
To make this expression concrete, we first examine the component variances.

The term $\langle (\Delta \hat{X}_{\text{coherent}})^2 \rangle$ represents the variance of the initial coherent state ($|\alpha\rangle$). Coherent states are fundamental in CV quantum systems, embodying minimum uncertainty states that saturate the Heisenberg uncertainty principle\cite{braunstein2005quantum, WANG2007Gaussian}. Their characteristic quadrature variance, for instance in the $\hat{X}$ quadrature, is the minimal shot noise variance\cite{braunstein2005quantum}: 
\begin{equation}
\langle (\Delta \hat{X}_{\text{coherent}})^2 \rangle = \frac{1}{2}.
\label{eq:coherent_variance}
\end{equation}
This value quantifies the minimal spread of the state's Gaussian profile in phase space\cite{braunstein2005quantum, wigner1932wigner}. 

The term $\langle (\Delta \hat{X}_{\text{thermal}})^2 \rangle$ pertains to the variance of the thermal states modeling the environment. These states are also Gaussian, and their variance increases with the average photon number $\bar{n}$\cite{WANG2007Gaussian}: 
\begin{equation}
\langle (\Delta \hat{X}_{\text{thermal}})^2 \rangle = \frac{1}{2} + \bar{n}.
\label{eq:thermal_variance}
\end{equation}
This increased variance reflects a broader and more mixed phase-space distribution compared to coherent or vacuum states. 

Substituting the specific expressions for coherent state variance (Equation \eqref{eq:coherent_variance}) and thermal state variance (Equation \eqref{eq:thermal_variance}) into the general combination formula (Equation \eqref{eq:total_variance_combination}), we derive the final explicit form for the total output variance:
\begin{equation}
\begin{aligned}
\langle (\Delta \hat{X})^2 \rangle_{\text{total}} &= \eta \cdot \frac{1}{2} + (1-\eta) \left( \frac{1}{2} + \bar{n} \right) \\
&= \frac{1}{2} + (1-\eta) \bar{n}.
\end{aligned}
\label{eq:total_variance_theory}
\end{equation}
This resulting equation links the output state's variance directly to the channel transmissivity $\eta$ and the environmental mean photon number $\bar{n}$ (for $\bar{n} > 0$). 
The term $(1-\eta)\bar{n}$ signifies the noise-induced broadening of the state's distribution and its increased mixedness\cite{WANG2007Gaussian, Holevo_2012}. 
This variance-noise relationship is key to our restoration strategy for thermally corrupted states.

\subsection{CVQD-R Training for Thermal Noise Restoration}
\label{subsubsec:cvqd_rm_training_new_label} 

In contrast to CVQD-G's objective of generating predefined target states from a noise prior, \textbf{CVQD-R is trained to restore arbitrary coherent states (within certain parameter ranges) that have been corrupted by thermal noise where $\bar{n} > 0$}. 

When $\bar{n} > 0$, Equation \eqref{eq:total_variance_theory} shows that the total variance $\langle (\Delta \hat{X})^2 \rangle_{\text{total}}$ depends on both $\eta$ and $\bar{n}$. 
If $\bar{n}$ is known or estimable, the received state's variance signals the noise level $\eta$, enabling the quantum circuit to recognize the corruption degree by analyzing this variance. 

CVQD-R is trained to restore any unknown coherent state $|\alpha\rangle$ (with amplitude $|\alpha| \leq D$, for some maximum $D$). 
Training employs a set of coherent states $\{|\beta\rangle\}$ (e.g., $|\beta| \leq D$). 
The model receives a diffused state $\rho_t^{(\beta)}$ (state $|\beta\rangle$ after $t$ diffusion steps) and aims to produce an estimate $\tilde{\rho}_{(t-1)}^{(\beta)}$ approximating the previous step's state $\rho_{(t-1)}^{(\beta)}$. 
The loss function for this step is 
\begin{equation}
L_{(t-1)} = 1 - F(\rho_{(t-1)}^{(\beta)}, \tilde{\rho}_{(t-1)}^{(\beta)}) + \gamma P(\tilde{\rho}_{(t-1)}^{(\beta)}),
\end{equation}
where $\beta$ is randomly sampled, $F$ is fidelity, and and $P$ is a regularization term that enforces the trace-one property ($Tr(\tilde{\rho}_{t-1})=1$), ensuring the physical validity of the predicted state during simulation.

\subsection{The Quantum State Restoration Process}
\label{subsubsec:cvqd_rm_process_new_label} 

Once trained, CVQD-R is employed to restore unknown, noise-corrupted coherent states. 
The restoration process begins by taking the received state as the initial input. From this starting point, the model iteratively applies its learned single-step denoising transformation. In each subsequent step of this backward denoising process, the output state from the preceding step serves as the input for the current step. This iterative refinement continues until the process reconstructs an estimate of the original, clean state $\tilde{\rho}_0$.
This operational approach for restoration shares similarities with the backward generation process of CVQD-G, yet they differ crucially in their initial inputs: while CVQD-R commences with the actual corrupted data, CVQD-G typically starts its generation from a standard noise state.





\section{Complexity Analysis}
\label{section:complex}

This section analyzes the computational complexity of our Continuous-Variable Quantum Diffusion Generative Model (CVQD-G) framework. We examine circuit depth complexity during inference, training computational complexity, and provide comparative analysis with existing continuous-variable quantum generative approaches.

\subsection{Circuit Depth Complexity}

The computational complexity of CVQD-G during inference is primarily determined by the number of diffusion steps $T$ and the circuit depth $L$ for each denoising network. For each timestep $t \in \{1, 2, \ldots, T\}$, we apply a parametrized quantum circuit with $L$ layers, resulting in an overall circuit depth complexity of:

\begin{equation}
\mathcal{O}(T \times L).
\end{equation}

Each layer comprises Gaussian operations (rotation, squeezing, and displacement gates) and beam-splitter interactions between qumodes. Our implementation of timestep embedding enables parameter sharing across diffusion steps, substantially reducing the total parameter count while preserving model performance.

\subsection{Training Complexity}

Our novel CVQD-G architecture introduces a significant computational advantage through a direct sampling approach. Unlike conventional diffusion models that require sequential processing through all previous timesteps, our method can generate the quantum state at any arbitrary timestep $t$ in a single circuit execution. This innovation fundamentally changes the training complexity.

For each iteration, we sample random timesteps for batch elements according to our training strategy, resulting in a training computational complexity of:

\begin{equation}
\mathcal{O}(I \times B \times L),
\end{equation}
where $I$ represents the number of training iterations, $B$ is the batch size, and $L$ is the circuit depth per step. Notably, this complexity no longer scales with the total number of diffusion timesteps $T$, providing a substantial computational advantage over traditional sequential diffusion approaches.

This direct sampling capability represents a key contribution of our work, enabling efficient training even with a large number of defined diffusion steps, as the training process accesses these timesteps in parallel rather than sequentially.

\subsection{Comparative Analysis}
Compared to other continuous-variable quantum generative approaches, CVQD-G offers improved computational efficiency. Traditional variational methods typically require separate encoding and decoding networks, effectively doubling circuit depth. In contrast, our diffusion-based approach utilizes a single denoising network applied iteratively.

Additionally, CVQD-G avoids the significant measurement and classical post-processing overhead associated with Optical GANs \cite{shrivastava2019optical}, where discriminator evaluations introduce substantial additional complexity during training. Table~\ref{tab:complexity_comparison} summarizes these comparative advantages.

\begin{table}[ht]
\centering
\caption{Complexity comparison between CVQD-G and other continuous-variable quantum generative models}
\label{tab:complexity_comparison}
\begin{tabular}{lccc}
\hline
\textbf{Model} & \textbf{Circuit Depth} & \textbf{Space} & \textbf{Measurement} \\
 & \textbf{Complexity} & \textbf{Complexity} & \textbf{Overhead} \\
\hline
CVQD-G & $\mathcal{O}(L)$ & $\mathcal{O}(1)$ & Low \\
Optical GANs \cite{shrivastava2019optical} & $\mathcal{O}(L_g + L_d)$ & $\mathcal{O}(1)$ & High \\

\hline
\end{tabular}
\end{table}

This complexity analysis provides crucial insights for hardware implementation and identifies optimization opportunities for resource-constrained quantum systems.

\section{Numerical Simulations}
\label{section:numerical_simulations}

This section rigorously evaluates the Continuous-Variable Quantum Diffusion Generative Model (CVQD-G) on two critical tasks: high-fidelity quantum state generation and robust quantum state recovery. Our experiments aim to demonstrate CVQD-G's generative versatility across diverse quantum states and its practical utility in noise mitigation, a key challenge in realistic quantum information processing.

\subsection{Experimental Setup}
\label{subsec:experimental_setup}

All numerical simulations were performed using Strawberry Fields \cite{Killoran2019strawberryfields,Bromley_2020} as the quantum circuit simulation engine, specifically leveraging its Fock backend \cite{killoran2018continuous}. This backend necessitates a finite cutoff dimension for the Fock basis representation. While essential for simulation, this cutoff is not a fundamental limitation of physical CV quantum devices, which naturally operate in an infinite-dimensional Hilbert space \cite{killoran2018continuous,Killoran2019strawberryfields}. Optimization of the CVQNN parameters within CVQD-G was conducted using TensorFlow \cite{abadi2016tensorflow} with the Adam optimizer \cite{kingma2014adam}, enhanced by an exponential learning rate decay schedule for stable and efficient convergence \cite{loshchilov2016sgdr}. The systematic search and fine-tuning of these optimization settings and other crucial model hyperparameters were facilitated using the Optuna hyperparameter optimization framework \cite{akiba2019optuna}. 

Table~\ref{tab:cvqdm_hyperparameters} summarizes the core hyperparameters employed for the majority of our experiments. Any specific adjustments made for particularly challenging state parameters (e.g., coherent states with large amplitudes) are noted in the relevant discussions.

\begin{table}[htbp]
\centering
\caption{General Hyperparameters for CVQD-G Experiments.}
\label{tab:cvqdm_hyperparameters}
\begin{tabular}{|l|c|}
\hline
\textbf{Hyperparameter} & \textbf{Value} \\ \hline
Cut-off Dimension & 15 \\ \hline
Layers (per diffusion step in CVQNN) & 30 \\ \hline
Batch Size & 24 \\ \hline
Epochs & 99 \\ \hline
Total Timesteps $T$ (Diffusion/Denoising) & 112 \\ \hline
Initial Noise Schedule $\beta_{\text{start}}$ & $1.0 \times 10^{-4}$ \\ \hline
Final Noise Schedule $\beta_{\text{end}}$ & $0.05$ \\ \hline
Initial Transmissivity $\eta_0$ (Forward Process) & 0.99974 \\ \hline
Final Transmissivity $\eta_T$ (Forward Process) & 0.99331 \\ \hline
Timestep Loss Weight $\lambda$ & $8.55 \times 10^{-5}$ \\ \hline
Initial Learning Rate & 0.00778 \\ \hline
Decay Steps (for learning rate) & 8 \\ \hline
Decay Rate (for learning rate) & 0.9427 \\ \hline
Normalization Penalty Weight $\gamma$ & 100 \\ \hline
\end{tabular}
\end{table}

\subsection{Quantum State Generation}
\label{subsec:state_generation}

We systematically evaluated CVQD-G's ability to generate a diverse set of target quantum states, including both Gaussian and non-Gaussian types, chosen for their relevance and varying complexity in CV quantum information.

\textbf{Evaluation Protocol:}
All generation experiments were conducted considering two distinct environmental noise conditions for the forward diffusion process:
\begin{enumerate}
    \item \textbf{Pure Loss Channel ($\bar{n}=0$):} The environment is a vacuum state (loss without added thermal photons).
    \item \textbf{Thermal Loss Channel ($\bar{n}=0.5$):} The environment is a thermal state (average photon number of 0.5), representing a more realistic noisy scenario.
\end{enumerate}
The forward diffusion transforms an initial target state $\rho_0$ into a significantly noisy state $\rho_T$. This process is visualized by the fidelity $F(\rho_0, \rho_t)$ (subplot (b) in Figures~\ref{fig:coh_alpha1_nbar0_combined} through \ref{fig:cat1_nbar0_combined}). It's important to note that $F(\rho_0, \rho_T)$ is generally non-zero, as $\rho_T$ (a thermal-like or highly mixed state) retains some overlap with $\rho_0$ within a finite Fock cutoff. The aim is to reach a standardized noisy state, effectively erasing most of $\rho_0$'s features, from which the reverse denoising can robustly begin.

Model robustness was further assessed by initiating the reverse denoising process from partially noise-corrupted states $\rho_p = \mathcal{E}(\rho_0,\eta)$ (where $\mathcal{E}$ is the thermal loss channel with average photon number $\bar{n}$). Subplot (c) in the aforementioned figures illustrates convergence from these varied noisy starting points.

\subsubsection{Gaussian State Generation}
\label{subsubsec:gaussian_generation}
Gaussian states are foundational in CV quantum information and are generally more accessible experimentally.

\begin{figure}[htbp]
\centering
\includegraphics[width=\linewidth]{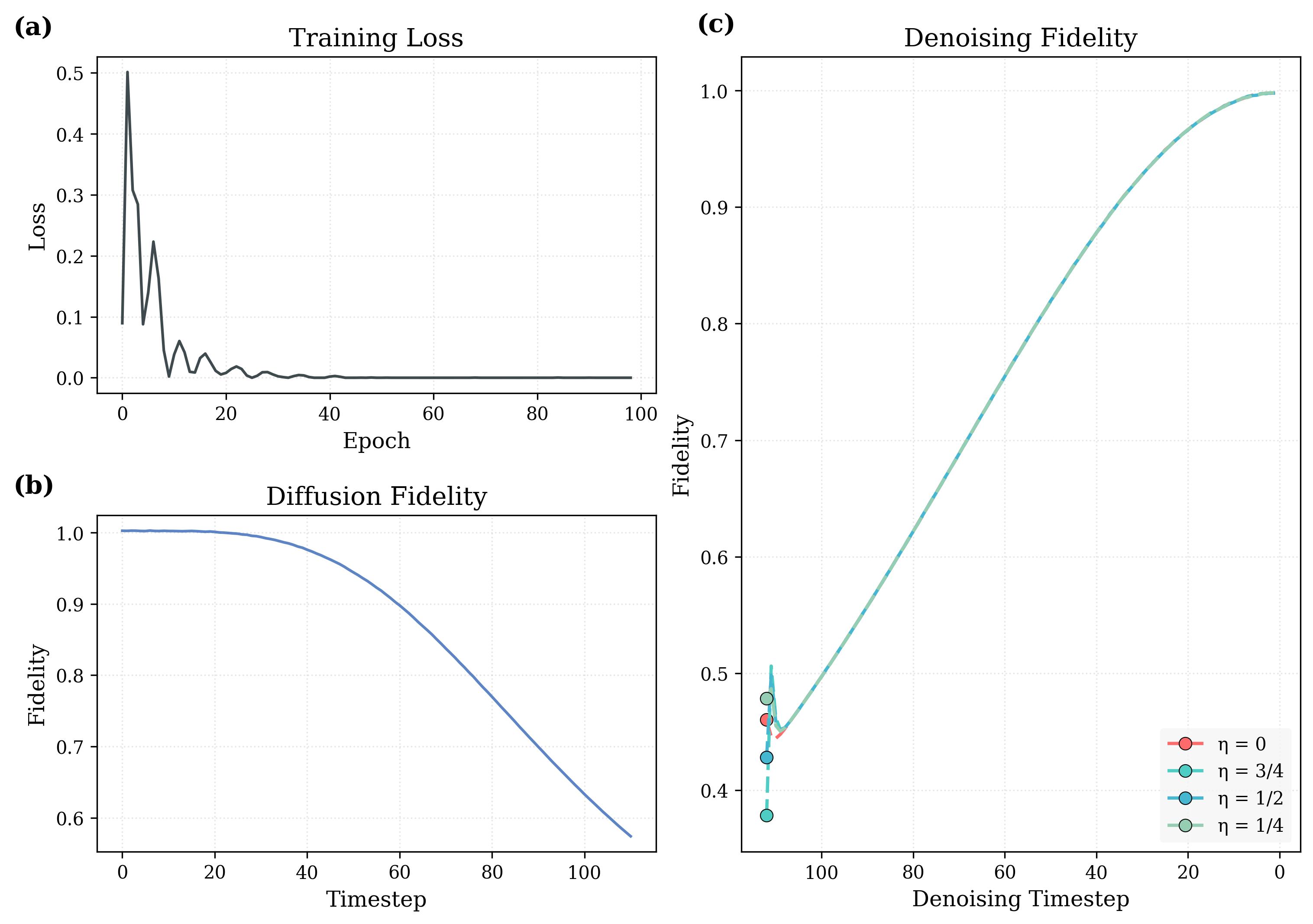} 
\caption{Generation of the coherent state $|\alpha=1.0\rangle$ in a pure loss channel ($\bar{n}=0$). (a) Training loss. (b) Forward diffusion fidelity. (c) Backward denoising fidelity from initial states with varying noise levels ($\eta$).}
\label{fig:coh_alpha1_nbar0_combined}
\end{figure}

\textbf{Coherent States ($|\alpha\rangle$):} These minimum uncertainty states, analogous to classical electromagnetic fields, are crucial for quantum communication. We first demonstrate generating $|\alpha = 1.0\rangle$. Figure~\ref{fig:coh_alpha1_nbar0_combined} details the training, forward diffusion, and backward denoising for this state in a pure loss ($\bar{n}=0$) environment, where CVQD-G achieved a high fidelity of 99.95\%. Subplot (c) confirms robust convergence to the target $|1.0\rangle$ even from initially noisy states (varied $\eta$). Similar high-fidelity generation was observed under thermal loss ($\bar{n}=0.5$), where CVQD-G attained 99.82\% for $|\alpha=1.0\rangle$. Detailed parameter sweep results for $\alpha \in \{0.5, 1.0, 1.5, 2.0, 2.5\}$ are shown in Figure~\ref{fig:param_sweeps_combined}(a) and detailed in Appendix~\ref{app:param_sweeps}. As $\alpha$ increases (e.g., $\alpha=2.5$), the finite Fock cutoff (dimension 15) impacts fidelity; however, targeted hyperparameter tuning (detailed in Appendix~\ref{app:hyperparam_tuning_coherent_alpha2.5}) can recover fidelities such as 97.80\% for $\bar{n}=0$ and 96.50\% for $\bar{n}=0.5$ with CVQD-G (detailed comparison of the general hyperparameter case in Appendix~\ref{app:param_sweeps}).

\begin{figure}[htbp]
\centering
\includegraphics[width=\linewidth]{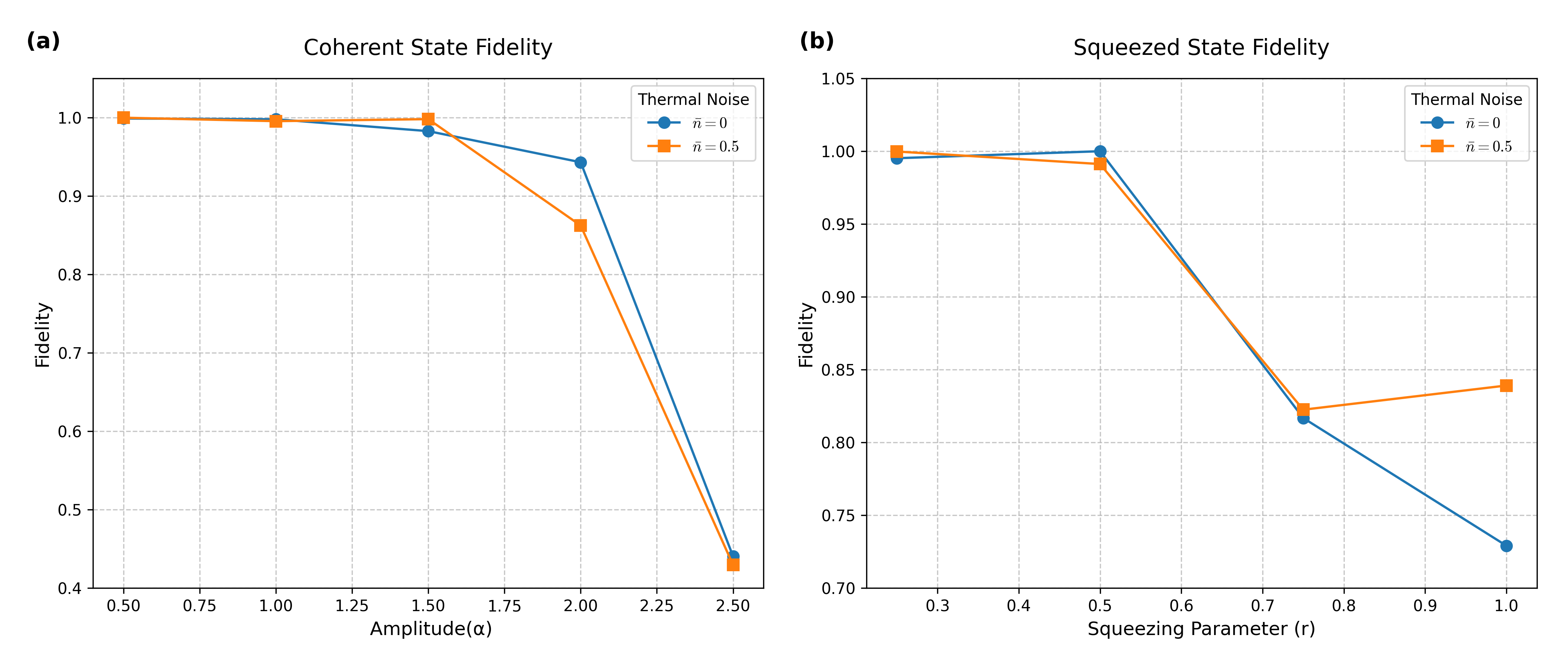} 
\caption{Fidelity of generated states as a function of key state parameters. (a) Coherent state fidelity versus amplitude $\alpha$. (b) Squeezed state fidelity versus squeezing parameter $r$. Both plots compare performance in pure loss ($\bar{n}=0$) and thermal loss ($\bar{n}=0.5$) environments, and the results were obtained using general hyperparameters.}
\label{fig:param_sweeps_combined}
\end{figure}

\begin{figure}[htbp]
\centering
\includegraphics[width=\linewidth]{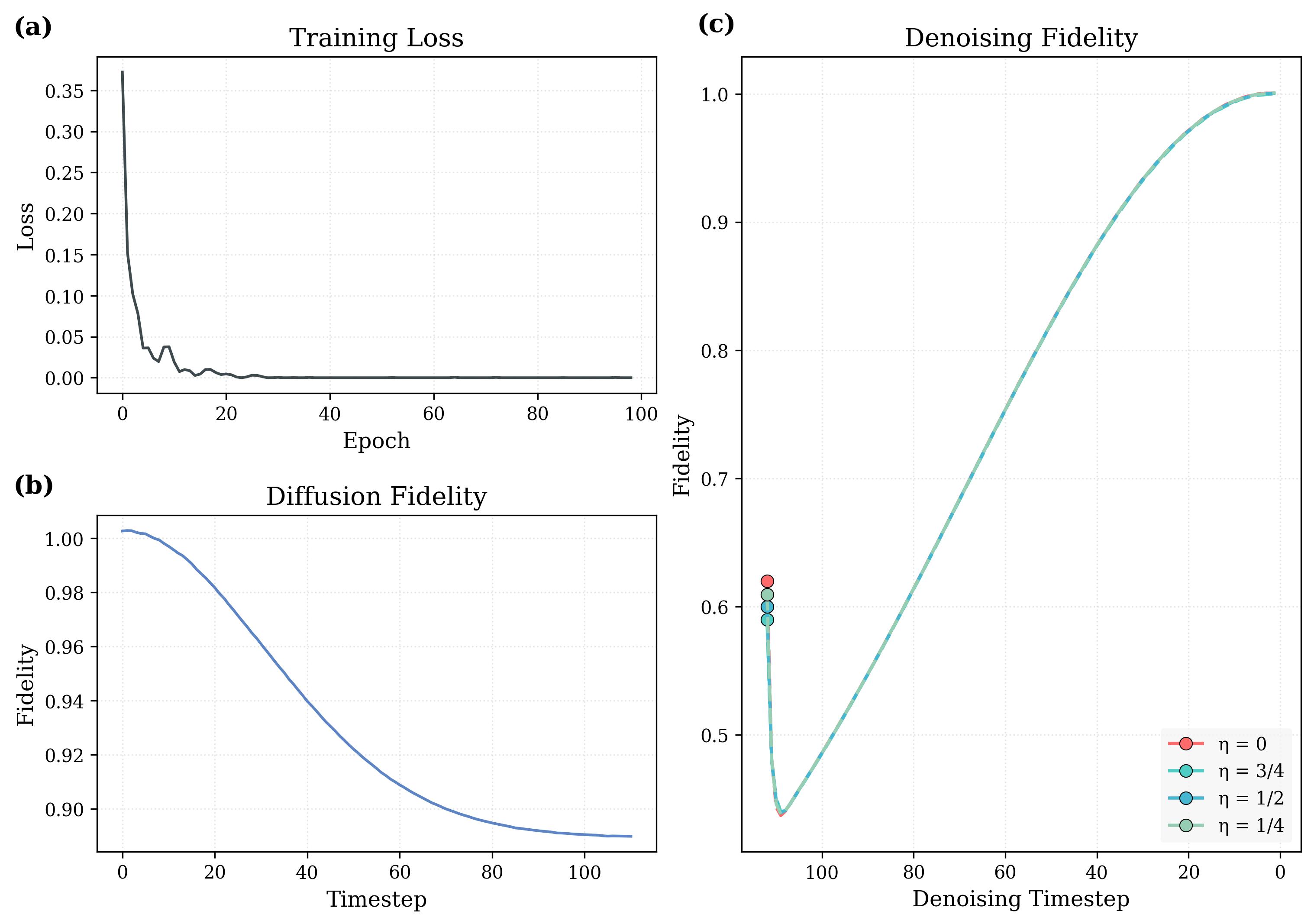}
\caption{Generation of the squeezed vacuum state $S(r=0.5)|0\rangle$ in a pure loss channel ($\bar{n}=0$). (a) Training loss. (b) Forward diffusion fidelity. (c) Backward denoising fidelity from initial states with varying noise levels ($\eta$).}
\label{fig:sq_r05_nbar0_combined}
\end{figure}

\textbf{Squeezed States ($S(r)|0\rangle$):} Squeezed vacuum states, featuring reduced quantum noise in one quadrature, are vital for high-precision measurements and as resources for complex state generation. We tested generation for a squeezing parameter $r=0.5$. Figure~\ref{fig:sq_r05_nbar0_combined} shows the process for $\bar{n}=0$, where CVQD-G yielded a high fidelity of 99.56\%. For $r=0.5$ under thermal loss ($\bar{n}=0.5$), CVQD-G achieved a fidelity of 99.03\%. Figure~\ref{fig:param_sweeps_combined}(b) shows the impact of varying $r \in \{0.25, 0.5, 0.75, 1.0\}$ on fidelity for both $\bar{n}=0$ and $\bar{n}=0.5$ environments using general hyperparameters. Increasing $r$ makes generation more challenging due to more pronounced non-classical features, leading to a slight fidelity decrease.

\subsubsection{Non-Gaussian State Generation}
\label{subsubsec:non_gaussian_generation}
Non-Gaussian states are critical for universal CV quantum computation and exhibit complex quantum features.

\begin{figure}[htbp]
\centering
\includegraphics[width=\linewidth]{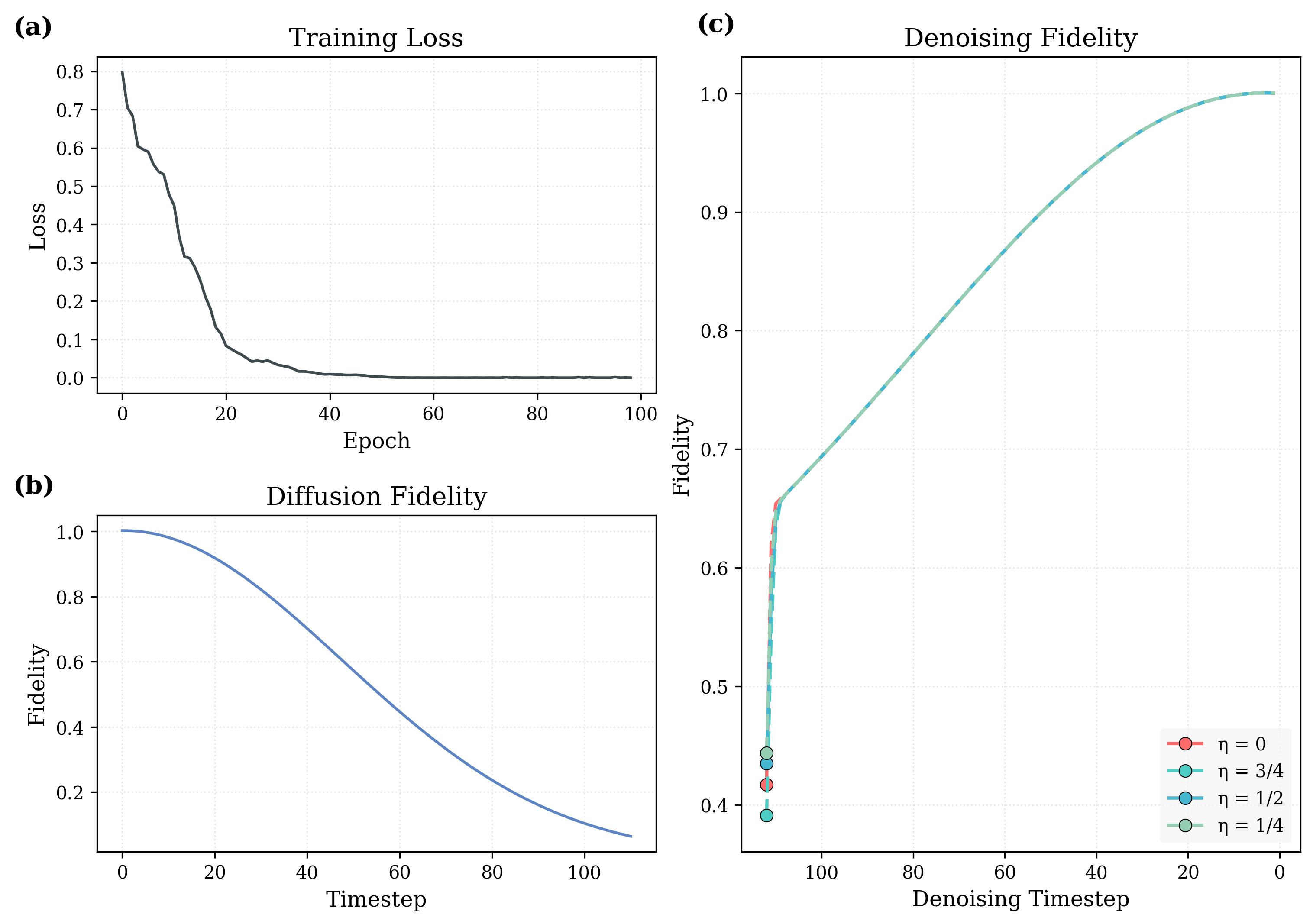}
\caption{Generation of the Fock state $|1\rangle$ in a pure loss channel ($\bar{n}=0$). (a) Training loss. (b) Forward diffusion fidelity. (c) Backward denoising fidelity from initial states with varying noise levels ($\eta$).}
\label{fig:fock1_nbar0_combined}
\end{figure}

\textbf{Fock States ($|n\rangle$):} These number states, with a precisely defined photon count, are fundamental non-classical resources. We demonstrate generating the first excited Fock state, $|1\rangle$. Figure~\ref{fig:fock1_nbar0_combined} shows results for $\bar{n}=0$, with CVQD-G achieving a fidelity of 99.85\%. Notably, the forward diffusion fidelity $F(\rho_0, \rho_t)$ for $|1\rangle$ (subplot (b)) approaches a very low value. This occurs because the distinct, non-classical nature of $|1\rangle$ has minimal overlap with the vacuum-like or thermal-like state $\rho_T$ targeted by the diffusion, effectively erasing its specific photon number characteristic. Similar high generation performance by CVQD-G is seen for $\bar{n}=0.5$ (detailed in Appendix~\ref{app:diverse_states}).

\begin{figure}[htbp]
\centering
\includegraphics[width=\linewidth]{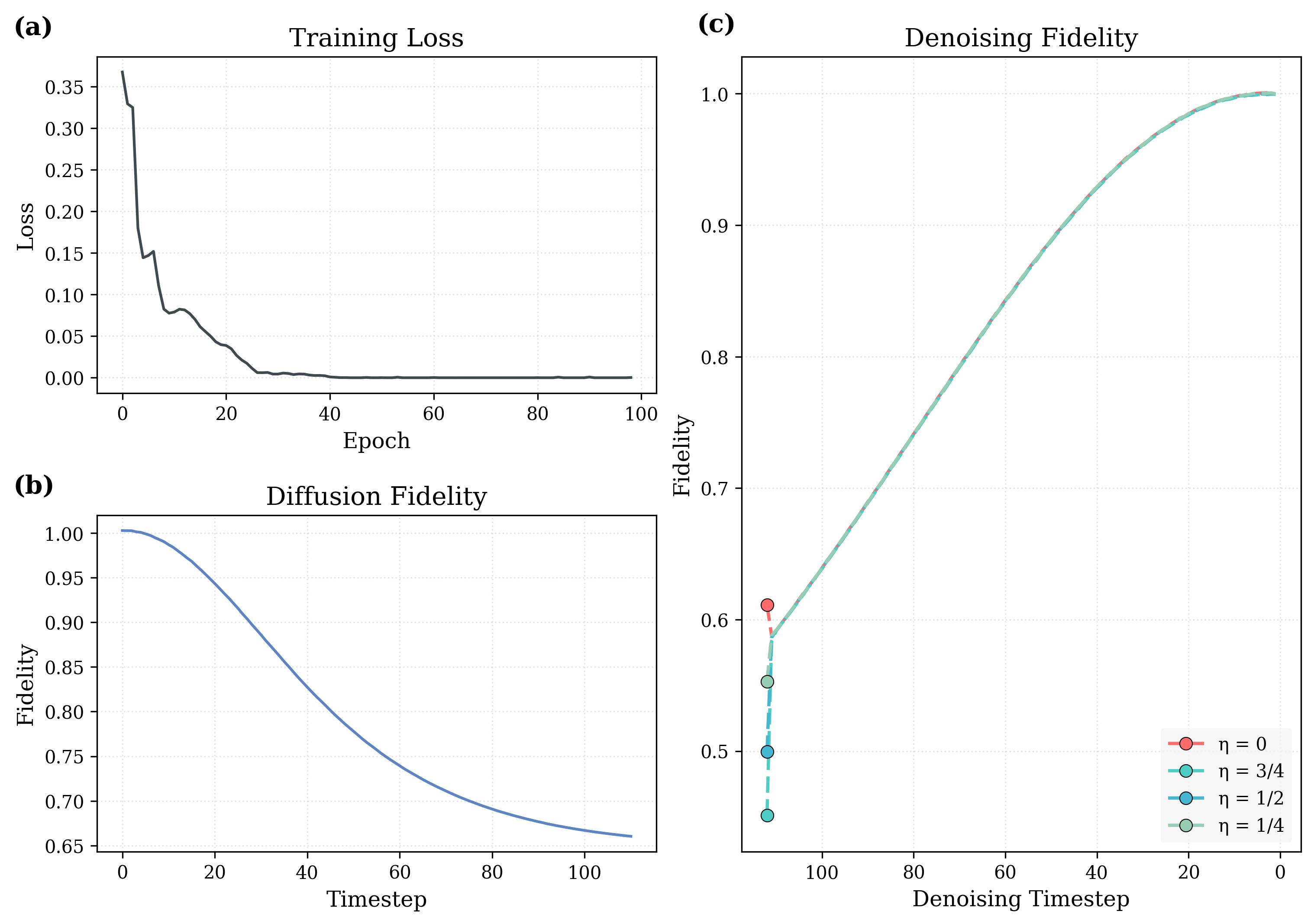}
\caption{Generation of the even cat state $|\text{cat}(1)\rangle$ in a pure loss channel ($\bar{n}=0$). (a) Training loss. (b) Forward diffusion fidelity. (c) Backward denoising fidelity from initial states with varying noise levels ($\eta$).}
\label{fig:cat1_nbar0_combined}
\end{figure}

\textbf{Cat States ($|\text{cat}(\alpha)\rangle$):} Cat states, superpositions of distinct coherent states (e.g., $|\text{cat}(\alpha)\rangle \propto (|\alpha\rangle \pm |-\alpha\rangle)$), embody macroscopic quantum superposition and are valuable for quantum error correction and metrology. We generated an even cat state $|\text{cat}(1)\rangle$. Figure~\ref{fig:cat1_nbar0_combined} shows generation in the $\bar{n}=0$ environment, where CVQD-G achieved a fidelity of 99.61\%. Cat states are generally more challenging to generate due to their delicate superposition. Performance for $\bar{n}=0.5$ is detailed in Appendix~\ref{app:diverse_states}.

\textbf{Comparative Performance with Optical GANs ($\mathbf{\bar{n}=0}$):}
For a direct comparison against Optical GANs \cite{shrivastava2019optical} under pure loss conditions ($\bar{n}=0$), Table~\ref{tab:gan_comparison_nbar0_fullwidth} summarizes the achieved fidelities. Our CVQD-G consistently demonstrates superior performance across all listed generated states. For instance, in generating the Coherent state $|\alpha=1.0\rangle$, CVQD-G achieved 99.95\% fidelity, surpassing the 98.50\% by Optical GANs. Similar advantages for CVQD-G are observed for Squeezed, Fock, and Cat states, as detailed in the table.

\begin{table*}[tp] 
\centering
\small 
\caption{Comparison of State Generation Fidelities (\%) for $\bar{n}=0$ Pure Loss Channels: CVQD-G vs. Optical GAN \cite{shrivastava2019optical}. Fidelities for CVQD-G are from this work; Optical GAN fidelities are for comparison from the cited source.}
\label{tab:gan_comparison_nbar0_fullwidth} 
\begin{tabular}{|l|c|c|c|c|}
\hline
\textbf{Model} & \makecell[c]{\textbf{Coherent State} \\ \textbf{($\alpha=1.0$)}} & \makecell[c]{\textbf{Squeezed State} \\ \textbf{($r=0.5$)}} & \makecell[c]{\textbf{Fock State} \\ \textbf{($n=1$)}} & \makecell[c]{\textbf{Cat State} \\ \textbf{($\alpha=1$)}} \\ \hline
CVQD-G (Ours) & \textbf{99.95} & \textbf{99.56} & \textbf{99.85} & \textbf{99.61} \\
Optical GAN \cite{shrivastava2019optical} & 83.43 & 94.95 & 73.27 & 78.82 \\ \hline
\end{tabular}
\end{table*}

\subsubsection{Key Observations from State Generation}
\label{subsubsec:generation_observations_revised}
The state generation experiments consistently highlight several key capabilities of CVQD-G:
\begin{enumerate}
    \item \textbf{Effective Learning of Noise Reversal:} CVQD-G consistently learns to reverse the forward diffusion, evidenced by high final fidelities for diverse target states. For $\bar{n}=0$ conditions, these results (Table~\ref{tab:gan_comparison_nbar0_fullwidth}) also show superior performance compared to Optical GAN benchmarks. The general performance is illustrated in Figures~\ref{fig:coh_alpha1_nbar0_combined}--\ref{fig:cat1_nbar0_combined} and Figure~\ref{fig:param_sweeps_combined}.
    \item \textbf{Robustness to Initial Conditions:} The model demonstrates strong convergence to the target state even when denoising is initiated from partially noise-corrupted states (subplot (c) in Figures~\ref{fig:coh_alpha1_nbar0_combined}--\ref{fig:cat1_nbar0_combined}).
    \item \textbf{Performance Across Noise Environments:} High generation fidelity is maintained by CVQD-G for Gaussian states in both pure loss ($\bar{n}=0$) and thermal loss ($\bar{n}=0.5$) environments (Figure~\ref{fig:param_sweeps_combined}). The specific CVQD-G fidelities for $\bar{n}=0.5$ (e.g., 99.82\% for Coherent $\alpha=1.0$; 96.50\% for tuned Coherent $\alpha=2.5$; 99.03\% for Squeezed $r=0.5$) confirm strong performance even with thermal noise. Similar robustness is noted for non-Gaussian states (see Appendix~\ref{app:diverse_states}).
    \item \textbf{Complexity vs. Fidelity Trade-off:} Achieving optimal fidelities for states with larger phase-space features (e.g., coherent states with large $\alpha$, for which tuned CVQD-G achieved 97.80\% for $\bar{n}=0$) or significant non-classicality may necessitate careful hyperparameter tuning (Appendix~\ref{app:hyperparam_tuning_coherent_alpha2.5}) or increased computational resources.
\end{enumerate}
These findings confirm CVQD-G's proficiency in learning the characteristics of quantum noise channels and effectively reversing their impact for generative tasks.

\subsection{Quantum State Restoration}
\label{subsec:state_restoration}

Beyond \textit{de novo} state generation, a critical application of our diffusion model framework is the restoration of quantum states that have been corrupted by noise. This section details the model's capability to restore coherent states from degradation caused by thermal loss channels, a scenario highly relevant for practical quantum communication. For these experiments, we utilize a model trained specifically for restoration tasks, which we can refer to as a Continuous-Variable Quantum Diffusion Restoration Model (CVQD-R) if distinct from the generative CVQD-G.

The training process for state restoration involves generating a diverse set of initial clean coherent states. These are created by applying an X-gate (displacement along the x-axis) with parameters randomly sampled from 0 to 1, combined with an R-gate (phase rotation) with parameters randomly sampled from 0 to $2\pi$. These clean states are then subjected to simulated noise by passing them through thermal loss channels with varying levels of corruption, corresponding to different timesteps in a diffusion process, similar to the training for state generation. The model is then trained to reverse these noise effects.

The core hyperparameters used for the state restoration experiments are summarized in Table~\ref{tab:CVQD-R_hyperparameters_restoration}. Parameters such as Cut-off Dimension and Layers (per diffusion step in CVQNN) remain consistent with those in Table~\ref{tab:cvqdm_hyperparameters} unless specified otherwise.

\begin{table}[htbp]
\centering
\caption{Hyperparameters for CVQD-R State Restoration Experiments.}
\label{tab:CVQD-R_hyperparameters_restoration}
\begin{tabular}{|l|c|}
\hline
\textbf{Hyperparameter} & \textbf{Value} \\ \hline
Cut-off Dimension & 15 \\ \hline
Layers (per diffusion step in CVQNN) & 30 \\ \hline
Batch Size & 48 \\ \hline
Epochs & 112 \\ \hline
Total Timesteps $T$ (Diffusion/Restoration) & 150 \\ \hline
Initial Noise Schedule $\beta_{\text{start}}$ & $1.0 \times 10^{-4}$ \\ \hline
Final Noise Schedule $\beta_{\text{end}}$ & $0.05$ \\ \hline
Initial Learning Rate & $0.00045$ \\ \hline
Decay Steps (for learning rate) & 24 \\ \hline
Decay Rate (for learning rate) & $0.906$ \\ \hline
Normalization Penalty Weight ($\gamma$) & 100 \\ \hline 
Timestep Loss Weight ($\lambda$) & 0.16 \\ \hline 
\end{tabular}
\end{table}

\subsubsection{Restoration in Thermal Loss Channels}
\label{subsubsec:thermal_loss_restoration}

Thermal loss channels introduce both signal attenuation and added thermal noise, making state restoration particularly challenging yet crucial for real-world applications. We demonstrate the model's restoration capabilities using an environment with $\bar{n}=0.5$ as a representative thermal noise level.

\textbf{Restoring Coherent States with Varying Amplitudes from Fixed Noise:}
We first assess the model's ability to restore various coherent states when the noise channel parameters are fixed (e.g., transmissivity $\eta=0.5$ and environmental $\bar{n}=0.5$). For this, initial clean coherent states were prepared using X-gate displacement parameters of $s \in \{0.3, 0.5, 0.7\}$. For each $s$, eight different phase states were created by applying an R-gate with rotation angles $k\pi/4$ (for $k = 0, 1, \dots, 7$), resulting in 24 distinct initial coherent states. Each of these states was then corrupted by the fixed thermal loss channel, and the model attempted to restore it.

\begin{figure}[htbp]
\centering
\includegraphics[width=0.8\linewidth]{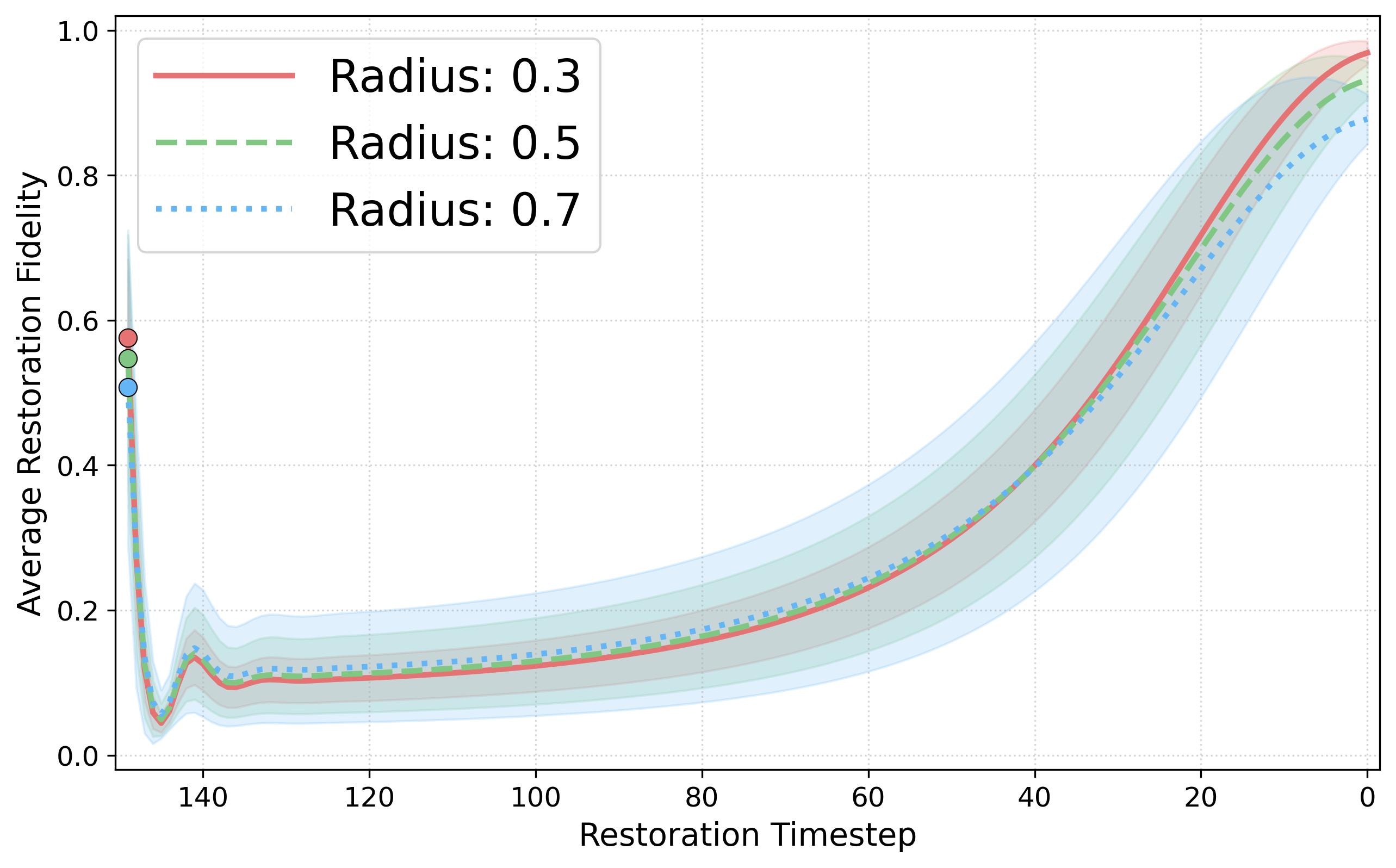} 
\caption{Average restoration fidelity versus restoration timestep for coherent states generated with varying X-gate parameters ($s=0.3, 0.5, 0.7$), averaged over 8 initial phases. States were corrupted by a fixed thermal loss channel (e.g., $\eta=0.5, \bar{n}=0.5$). Shaded regions might indicate standard deviation over phases.}
\label{fig:restore_varying_alpha_fixed_noise}
\end{figure}

Figure~\ref{fig:restore_varying_alpha_fixed_noise} illustrates the average restoration fidelity for these states, grouped by the initial X-gate parameter $s$. The curves show the fidelity of the restored state with the original clean state as a function of the restoration timestep. The model achieves high final restoration fidelities, specifically approximately 98\% for initial states with X-gate parameter $s=0.3$, 96\% for $s=0.5$, and 89\% for $s=0.7$. These results demonstrate the model's effectiveness in recovering states of different initial amplitudes when subjected to a consistent noise environment.

\textbf{Restoring a Fixed Coherent State from Variable Noise Levels:}
Another critical scenario involves restoring a specific, known initial coherent state that has been subjected to thermal loss channels with varying degrees of noise. For this test, the initial state was prepared using an X-gate parameter of $s=0.5$ and an R-gate parameter of $\pi/4$. This state was then passed through thermal loss channels with different effective transmissivities $\eta$ (corresponding to BS parameters $\pi/3, \pi/4, \pi/6$, resulting in $\eta$ values of $0.25, 0.50, \text{and } 0.75$ respectively, with environmental $\bar{n}=0.5$).

\begin{figure}[htbp]
\centering
\includegraphics[width=0.8\linewidth]{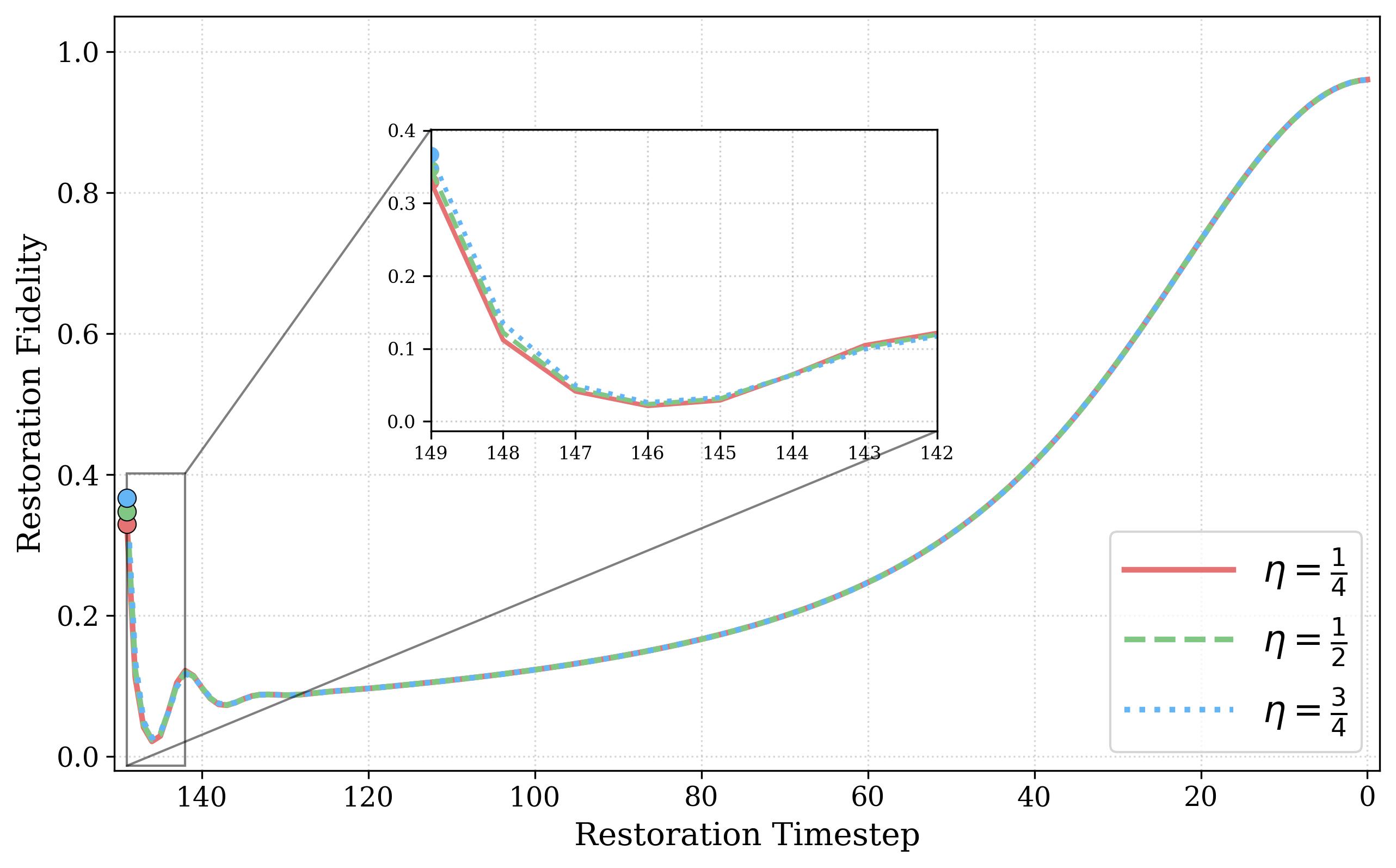} 
\caption{Restoration fidelity versus restoration timestep for a fixed initial coherent state (X-gate parameter $s=0.5$, R-gate parameter $\pi/4$) after corruption by thermal loss channels with varying effective transmissivities $\eta$ (e.g., $\eta \in \{0.25, 0.50, 0.75\}$; environmental $\bar{n}=0.5$).}
\label{fig:restore_fixed_alpha_variable_noise}
\end{figure}

Figure~\ref{fig:restore_fixed_alpha_variable_noise} shows the restoration fidelity curves for this fixed initial state when recovering from these different noise levels. Each curve represents the fidelity with the original clean state as a function of the restoration timestep for a given initial channel transmissivity $\eta$. A remarkable finding from this result is that the restoration fidelity curves for all tested channel transmissivities ($\eta \in \{0.25, 0.50, 0.75\}$) are nearly identical. This demonstrates that the model consistently restores the fixed initial state to a high fidelity of approximately 96\%. More importantly, the coincidence of these trajectories suggests that the restoration process is largely independent of the initial degree of noise corruption, underscoring the model's robustness and its ability to learn a generalized restoration mapping.

\subsection{Simulation Conclusions and Comparative Remarks}
\label{subsec:simulation_conclusion}

The comprehensive numerical simulations presented in this section robustly demonstrate the CVQD-G's effectiveness and versatility. Key takeaways include:
\begin{enumerate}
    \item \textbf{High-Fidelity State Generation:} CVQD-G successfully generates a wide array of both Gaussian (coherent, squeezed, as shown in Figures~\ref{fig:coh_alpha1_nbar0_combined}, \ref{fig:sq_r05_nbar0_combined}, \ref{fig:param_sweeps_combined}) and non-Gaussian (Fock, cat, as shown in Figures~\ref{fig:fock1_nbar0_combined}, \ref{fig:cat1_nbar0_combined}) states with high fidelities across different environmental noise conditions ($\bar{n}=0$ and $\bar{n}=0.5$).
    \item \textbf{Robust State Restoration:} The model exhibits strong capabilities in restoring coherent states corrupted by thermal loss channels ($\bar{n}=0.5$). It achieves high fidelities when restoring states of varying initial amplitudes from fixed noise (approx. 89\%-98\% for $s \in \{0.3,0.5,0.7\}$ in Figure~\ref{fig:restore_varying_alpha_fixed_noise}) and consistently restores a fixed state to high fidelity (approx. 96\%) from varying initial noise levels (Figure~\ref{fig:restore_fixed_alpha_variable_noise}).
    \item \textbf{Adaptability and Learned Generalization:} The model demonstrates remarkable robustness. In generation, its performance is resilient to variations in initial state noise (Figures~\ref{fig:coh_alpha1_nbar0_combined}(c)-\ref{fig:cat1_nbar0_combined}(c)) and channel noise characteristics (Figure~\ref{fig:param_sweeps_combined}). Crucially, in restoration, the near-identical fidelity curves for varied initial noise levels (Figure~\ref{fig:restore_fixed_alpha_variable_noise}) suggest the model learns a generalized restoration mapping, rather than distinct paths for each noise level. This indicates that the restoration trajectory is governed by a learned "field" largely independent of the initial corruption severity, underpinning the model's adaptability to unpredictable noise.
\end{enumerate}
These capabilities position CVQD-G (and its specialized variant for denoising) as a powerful and promising tool for advanced quantum state engineering and for mitigating noise in quantum communication protocols.

Compared to other quantum machine learning approaches for CV systems, such as Optical GANs \cite{shrivastava2019optical}, CVQD-G offers distinct advantages. While Optical GANs have shown promise in state generation, they typically rely on a generator-discriminator architecture which can be complex to train and may involve additional overhead from measurements if the discriminator is classical or requires quantum state comparisons. CVQD-G, by contrast:
\begin{itemize}
    \item Implements a physically motivated forward diffusion process (thermal loss), making the learned reverse process directly relevant to counteracting realistic noise.
    \item Does not require an adversarial discriminator, potentially simplifying the training dynamics and resource requirements.
    \item Is inherently designed for iterative refinement, which naturally lends itself to both generation (building complexity from noise) and restoration (incrementally removing noise). The step-wise denoising is particularly well-suited for handling varying noise levels, a challenge for models with more rigid input-output mappings.
    \item As indicated by our complexity analysis (Section~\ref{subsec:contributions} referring to $O(I \times B \times L)$ training complexity), CVQD-G can be efficient, especially due to parameter sharing across timesteps.
\end{itemize}
While a direct, exhaustive benchmark against other continuous-variable (CV) quantum generative models is beyond the scope of this initial work, the architectural differences and the demonstrated dual capability for robust generation and restoration suggest that CVQD-G provides a flexible and powerful alternative, particularly for applications demanding noise resilience and state recovery.

Additional experimental details, results for other quantum states generated (e.g. detailed sweeps for Gaussian states and non-Gaussian states under thermal loss), are provided in Appendix~\ref{app:param_sweeps} and Appendix~\ref{app:diverse_states}. Further details on hyperparameter tuning for specific cases can be found in Appendix~\ref{app:hyperparam_tuning_coherent_alpha2.5}.

\section{Conclusion}
\label{section: conclusion}
Inspired by classical diffusion generative models, this paper introduces the Continuous-Variable Quantum Diffusion Generative Model (CVQD-G), a novel approach for generating continuous-variable quantum states and restoring thermally polluted quantum states. The fundamental idea behind CVQD-G is that any target quantum state can be degraded into a thermal state, and a trainable backward process can reconstruct the original state.

In the forward process, a timestep-dependent thermal loss channel transforms the target state into a thermal state. In the backward process, a combination of a time-embedding quantum circuit and a denoising continuous-variable quantum neural network (CVQNN) is employed. The CVQNN shares the same parameters across different timesteps, while the time-embedding quantum circuit encodes timestep information into the quantum state, enabling the CVQNN to process the quantum state and generate a less noisy version.

To further specialize the quantum state restoration capability of CVQD-G, we propose the Continuous-Variable Quantum Diffusion Restoration Model (CVQD-R), an extension of CVQD-G designed to restore coherent states from an unknown level of thermal noise. Unlike CVQD-G, which aims to generate a target state, CVQD-R is trained on a set of states, allowing it to restore thermally polluted states even when the original state and noise level are unknown. This property makes CVQD-R more suitable for real-world applications, such as quantum communication.

Our work introduces a new model for continuous-variable quantum state generation. These models overcome the limitations of CVQNN-based quantum state generation, which relies on a fixed mapping from an initial state \( |\psi_0\rangle \) to a target state \( |\psi_t\rangle \). Instead, our approach provides a more flexible quantum state generation method with fewer constraints on the initial state. Furthermore, this model extends its capabilities beyond quantum state generation to quantum denoising, offering a new paradigm for quantum diffusion models.

Future research should explore state generation and denoising under a broader range of noise models. Additionally, optimizing quantum circuits to reduce the number of gates in the denoising process could improve resource efficiency. Another key challenge is that CVQD-G currently requires dedicated hyperparameters for different target states. Future studies could focus on optimizing algorithms and circuit designs to enable a universal set of hyperparameters applicable to various quantum states, thereby improving practical efficiency in real-world applications.


%

\appendices


\ifCLASSOPTIONcaptionsoff
  \newpage
\fi

\bibliographystyle{IEEEtran} 
\bibliography{references}  

\newpage
\onecolumn



\section{Proof of Theorem 1} 
\label{app:theorem1_proof}

We proceed with the proof in two parts. First, we derive the transformation in the Heisenberg picture by recursive application of the thermal loss channel, and then we establish the equivalent representation in the Schrödinger picture.

\medskip

\noindent\textit{Part 1: Heisenberg Picture Derivation}

\medskip
We begin by examining how the annihilation operator transforms under a single thermal loss channel at timestep $i$:
\begin{equation}
\hat{a}_i = \sqrt{\eta_i} \, \hat{a}_{i-1} + \sqrt{1 - \eta_i} \, \hat{e}_i,
\end{equation}
where each $\hat{e}_i$ represents an environmental mode in a thermal state with mean photon number $\bar{n}$.

Applying this transformation recursively for the first two steps:
\begin{align}
\hat{a}_1 &= \sqrt{\eta_1} \, \hat{a}_0 + \sqrt{1 - \eta_1} \, \hat{e}_1\\
\hat{a}_2 &= \sqrt{\eta_2} \, \hat{a}_1 + \sqrt{1 - \eta_2} \, \hat{e}_2\\
&= \sqrt{\eta_1\eta_2} \, \hat{a}_0 + \sqrt{\eta_2(1 - \eta_1)} \, \hat{e}_1 + \sqrt{1 - \eta_2} \, \hat{e}_2.
\end{align}

To simplify this expression, we define an effective environmental operator $\hat{E}_2$ such that:
\begin{equation}
\sqrt{1-\eta_1\eta_2}\hat{E}_2 = \sqrt{\eta_2(1-\eta_1)}\hat{e}_1 + \sqrt{1-\eta_2}\hat{e}_2.
\end{equation}

We can verify that $\hat{E}_2$ is a proper bosonic operator satisfying the commutation relation $[\hat{E}_2,\hat{E}_2^\dagger]=1$:
\begin{align}
[\hat{E}_2,\hat{E}_2^\dagger] &= \frac{1}{1-\eta_1\eta_2}\left[\eta_2(1-\eta_1)[\hat{e}_1,\hat{e}_1^\dagger] + (1-\eta_2)[\hat{e}_2,\hat{e}_2^\dagger]\right]\\
&= \frac{\eta_2(1-\eta_1) + (1-\eta_2)}{1-\eta_1\eta_2} = \frac{1-\eta_1\eta_2}{1-\eta_1\eta_2} = 1.
\end{align}

Moreover, when all environmental modes have the same mean photon number $\bar{n}$, $\hat{E}_2$ also maintains this mean photon number:
\begin{align}
\langle\hat{E}_2^\dagger\hat{E}_2\rangle &= \frac{\eta_2(1-\eta_1)\langle\hat{e}_1^\dagger\hat{e}_1\rangle + (1-\eta_2)\langle\hat{e}_2^\dagger\hat{e}_2\rangle}{1-\eta_1\eta_2}\\
&= \frac{\bar{n}[\eta_2(1-\eta_1) + (1-\eta_2)]}{1-\eta_1\eta_2} = \bar{n}.
\end{align}

This shows that $\hat{E}_2$ corresponds to a thermal state with the same mean photon number $\bar{n}$ as the original environmental modes. Therefore, we can express $\hat{a}_2$ compactly as:
\begin{equation}
\hat{a}_2 = \sqrt{\eta_1\eta_2}\hat{a}_0 + \sqrt{1-\eta_1\eta_2}\hat{E}_2.
\end{equation}

By induction, using $\bar{\eta}_t = \prod_{i=1}^{t} \eta_i$, we arrive at the general form for any timestep $t$:
\begin{equation}
\hat{a}_t = \sqrt{\bar{\eta}_t}\hat{a}_0 + \sqrt{1-\bar{\eta}_t}\hat{E}_t,
\end{equation}
where $\hat{E}_t$ remains a thermal environmental mode with mean photon number $\bar{n}$.
\medskip

\noindent\textit{Part 2: Schrödinger Picture Representation}
\medskip

To transform into the Schrödinger picture, we recognize that this transformation is precisely what occurs when applying a beam splitter operation between the system and a thermal environment. According to the beam splitter transformation:
\begin{equation}
U_{\text{BS}}(\eta): \begin{pmatrix} \hat{a} \\ \hat{b} \end{pmatrix} \mapsto \begin{pmatrix} \sqrt{\eta}\hat{a} + \sqrt{1-\eta}\hat{b} \\ -\sqrt{1-\eta}\hat{a} + \sqrt{\eta}\hat{b} \end{pmatrix}.
\end{equation}

When we apply this with transmissivity $\bar{\eta}_t$ to the initial system mode $\hat{a}_0$ and an environmental mode $\hat{E}$ in a thermal state $\rho_{\text{th}}(\bar{n})$, the system mode transforms exactly as our derived expression above. In the Schrödinger picture, this corresponds to:
\begin{equation}
\rho_t = \mathrm{Tr}_{\text{E}}[U_{\text{BS}}(\bar{\eta}_t)(\rho_{0} \otimes \rho_{\text{th}}(\bar{n}))U_{\text{BS}}^{\dagger}(\bar{\eta}_t)].
\end{equation}

Therefore, we can directly obtain $\rho_t$ from $\rho_0$ in a single step using one thermal loss channel with transmissivity $\bar{\eta}_t = \prod_{i=1}^{t} \eta_i$, rather than requiring $t$ sequential applications.
\hfill $\square$

\clearpage 

\noindent For the subsequent appendices (\ref{app:param_sweeps}, \ref{app:diverse_states}, and \ref{app:hyperparam_tuning_coherent_alpha2.5}) detailing numerical simulation results, the following plot conventions apply unless otherwise specified: plots showing training loss use an orange solid line, forward diffusion fidelity uses a red dashed line, and backward denoising fidelity uses a blue solid line.


\section{Quantum State Generation with Varying State Parameters} 
\label{app:param_sweeps}

This section details the CVQD-G's performance for generating Gaussian states with varying characteristic parameters using the general hyperparameters listed in Table~\ref{tab:cvqdm_hyperparameters} (main text) and a pure loss ($\bar{n}=0$) environment.

\subsection{Coherent States ($|\alpha\rangle$) vs. Amplitude $\alpha$}
\begin{figure}[htbp]
\centering
\includegraphics[width=\linewidth]{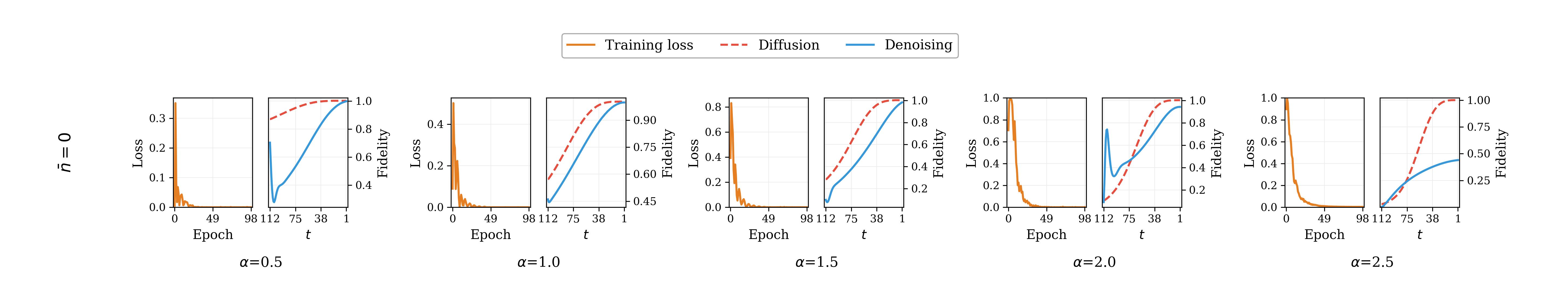}
\caption{Coherent state ($|\alpha\rangle$) generation with $\bar{n}=0$ for amplitudes $\alpha=0.5, 1.0, 1.5, 2.0, \text{and } 2.5$ (left to right), using general hyperparameters. Each pair shows training loss (left) and fidelities (right).}
\label{fig:app_coherent_nbar0_sweep_appendix}
\end{figure}

Figure~\ref{fig:app_coherent_nbar0_sweep_appendix} shows the generation results for coherent states $|\alpha\rangle$ with amplitudes $\alpha \in \{0.5, 1.0, 1.5, 2.0, 2.5\}$. For $\alpha=2.5$, the fidelity achieved with general hyperparameters is affected by the fixed Fock cutoff dimension, as discussed in Section~\ref{subsec:state_generation}. For results with tuned hyperparameters for $\alpha=2.5$, see Appendix~\ref{app:hyperparam_tuning_coherent_alpha2.5}.

\subsection{Squeezed States ($S(r)|0\rangle$) vs. Squeezing Parameter $r$}
\begin{figure}[htbp]
\centering
\includegraphics[width=\linewidth]{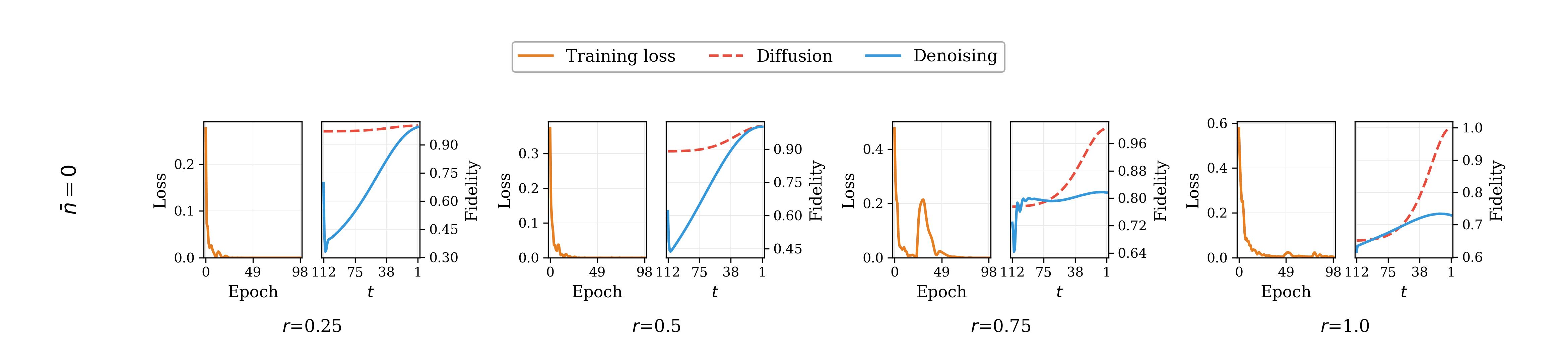}
\caption{Squeezed vacuum state ($S(r)|0\rangle$) generation with $\bar{n}=0$ for squeezing $r=0.25, 0.5, 0.75, \text{and } 1.0$ (left to right), using general hyperparameters. Each pair shows training loss (left) and fidelities (right).}
\label{fig:app_squeezed_nbar0_sweep_appendix}
\end{figure}

Figure~\ref{fig:app_squeezed_nbar0_sweep_appendix} presents results for squeezed vacuum states $S(r)|0\rangle$ with squeezing parameters $r \in \{0.25, 0.5, 0.75, 1.0\}$ using general hyperparameters. A slight fidelity decrease is observed for larger $r$ values.

\section{Generation of Diverse Quantum States} 
\label{app:diverse_states}

This section summarizes the CVQD-G's performance in generating a variety of quantum states (Coherent, Squeezed, Fock, and Cat states) under both pure loss ($\bar{n}=0$) and thermal loss ($\bar{n}=0.5$) environments, using the general hyperparameters from Table~\ref{tab:cvqdm_hyperparameters}. These results support discussions in Sections~\ref{subsec:state_generation} and \ref{subsubsec:non_gaussian_generation}.

\begin{figure}[htbp]
\centering
\includegraphics[width=\linewidth]{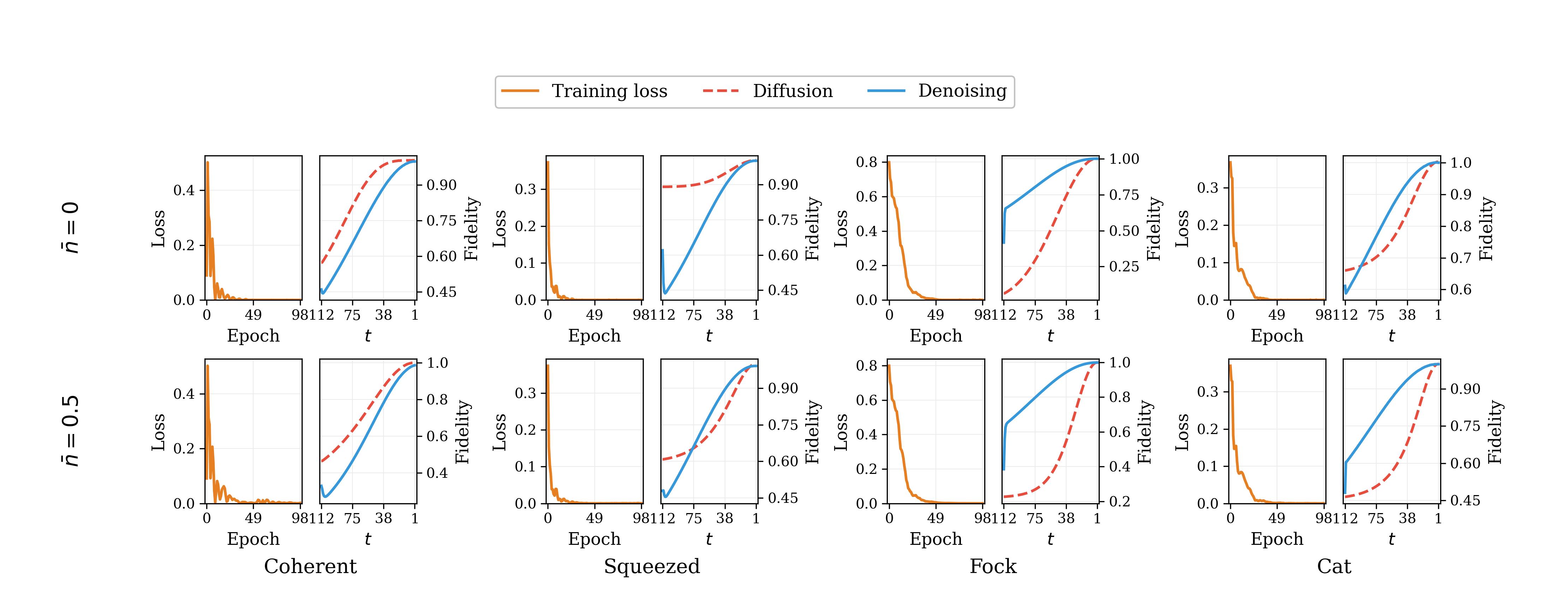}
\caption{Summary of training loss and fidelities. Top row: $\bar{n}=0$ (pure loss). Bottom row: $\bar{n}=0.5$ (thermal loss). States (left to right): Coherent ($|\alpha=1.0\rangle$), Squeezed ($S(r=0.5)|0\rangle$), Fock ($|1\rangle$), and Cat ($|\text{cat}(1)\rangle$).}
\label{fig:app_summary_generation_appendix}
\end{figure}

Figure~\ref{fig:app_summary_generation_appendix} illustrates that high fidelities are achieved for these diverse states.
For Gaussian states (Coherent, Squeezed) under thermal loss ($\bar{n}=0.5$), the model maintains robust performance. For specific parameter sweep fidelities under thermal loss, refer to Figure~\ref{fig:param_sweeps_combined} in the main text.
For non-Gaussian states (Fock, Cat), high fidelities are obtained in pure loss conditions (e.g., \textgreater99.8\% for Fock $|1\rangle$, and ~99.6\% for Cat $|\text{cat}(1)\rangle$, as noted in the main text). Performance remains strong under thermal noise. The characteristic low forward diffusion fidelity for Fock states is also evident.

\section{Targeted Hyperparameter Optimization for Coherent State $|\alpha=2.5\rangle$} 
\label{app:hyperparam_tuning_coherent_alpha2.5}

As mentioned in Section~\ref{subsec:state_generation}, generating coherent states with large amplitudes, such as $|\alpha=2.5\rangle$, can be challenging with general hyperparameters due to the fixed Fock cutoff dimension (15). To achieve higher fidelity for $|\alpha=2.5\rangle$ in a pure loss ($\bar{n}=0$) environment, a targeted set of hyperparameters was employed. These specialized hyperparameters, differing from or matching the general parameters in Table~\ref{tab:cvqdm_hyperparameters} as indicated, are summarized in Table~\ref{tab:tuned_hyperparams_coherent_alpha2.5}.

\begin{table}[htbp]
\centering
\caption{Specialized Hyperparameters for Coherent State $|\alpha=2.5\rangle$ ($\bar{n}=0$).}
\label{tab:tuned_hyperparams_coherent_alpha2.5}
\begin{tabular}{|l|c|}
\hline
\textbf{Hyperparameter} & \textbf{Value} \\ \hline
Batch Size & 30 \\ \hline
Epochs & 78 \\ \hline
Total Timesteps $T$ & 117 \\ \hline
Initial Learning Rate & $0.002605$ \\ \hline
Decay Steps (learning rate) & 15 \\ \hline
Decay Rate (learning rate) & $0.8394$ \\ \hline
Timestep Loss Weight $\lambda$ & $2.3236 \times 10^{-5}$ \\ \hline
Normalization Penalty Weight $\gamma$ & $10.358$ \\ \hline
Initial Noise Schedule $\beta_{\text{start}}$ & $1.0 \times 10^{-4}$ (as general) \\ \hline
Final Noise Schedule $\beta_{\text{end}}$ & $0.05$ (as general) \\ \hline
\end{tabular}
\end{table}

\begin{figure}[ht] 
\centering
\includegraphics[width=0.4\linewidth]{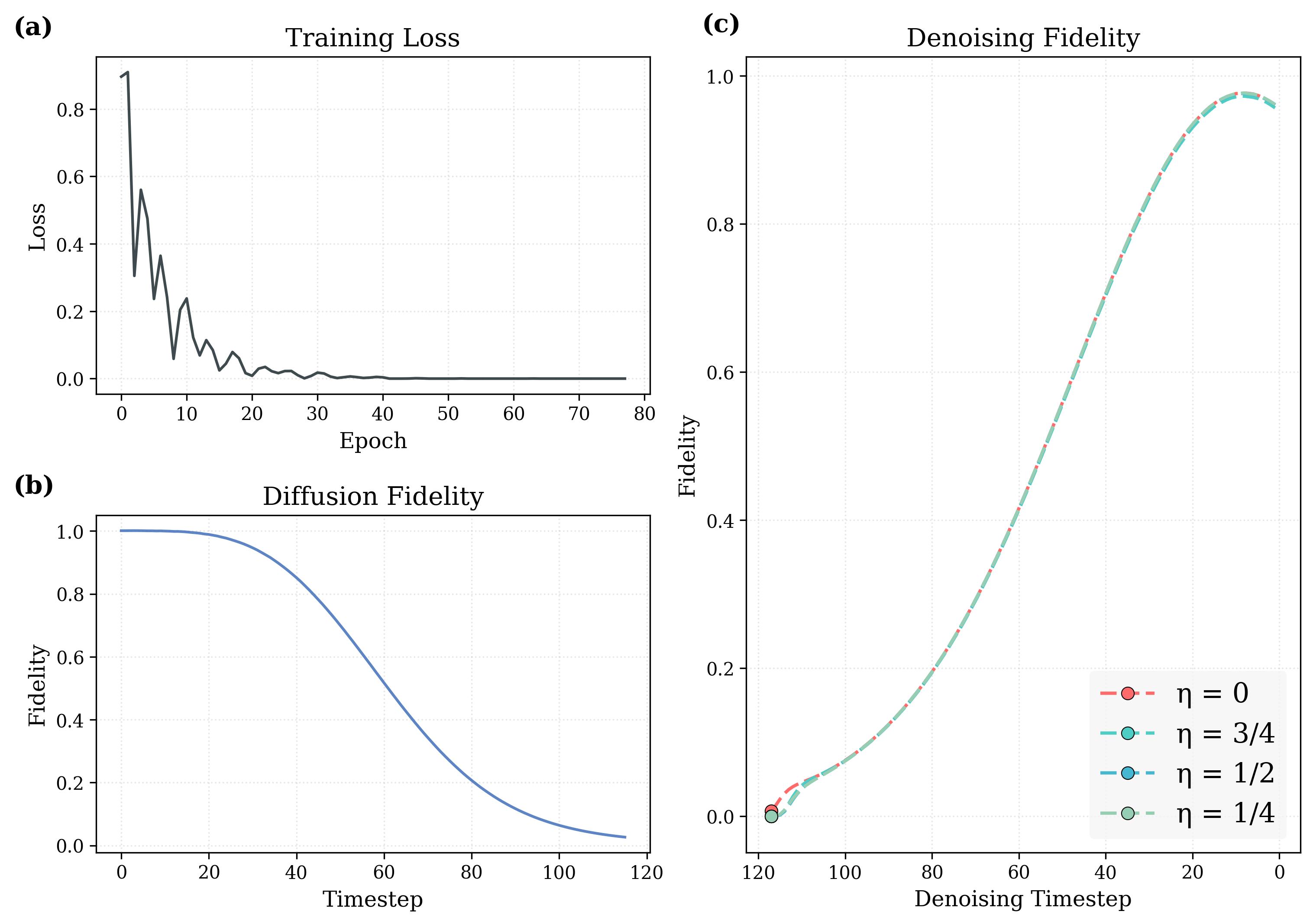} 
\caption{Performance for coherent state $|\alpha=2.5\rangle$ with $\bar{n}=0$ using tuned hyperparameters. (a) Training loss vs. epoch. (b) Forward diffusion fidelity $F(\rho_0, \rho_t)$ vs. timestep. (c) Backward denoising fidelity $F(\rho_{\text{0}}, \rho_{t})$ vs. denoising timestep, achieving \textgreater97\% fidelity.} 
\label{fig:app_coherent_alpha25_tuned_appendix}
\end{figure}

Figure~\ref{fig:app_coherent_alpha25_tuned_appendix} shows the training loss, forward diffusion fidelity, and backward denoising fidelity for the coherent state $|\alpha=2.5\rangle$ generated using these tuned hyperparameters, achieving a final fidelity exceeding 97\%.

%
\end{document}